\documentclass[aps,pra,twocolumn,showpacs,amssymb,amsmath,amsfonts,superscriptaddress,floatfix,nofootinbib,groupedaddress]
{revtex4-2}
\usepackage{graphicx}
\usepackage{mathtools}
\usepackage{perpage} 
\usepackage{amsmath,amssymb}
\MakePerPage{footnote}
\usepackage{color}
\usepackage{tabularx}
\usepackage{braket}
\usepackage{subfigure}
\usepackage{amsmath}
\usepackage{amsmath, amssymb}
\usepackage{algorithm}
\usepackage{algpseudocode}
\usepackage{fullpage}
\usepackage{xcolor}

\begin{document}


\title{Development of Neural Network-Based Optimal Control Pulse Generator for Quantum Logic Gates Using the GRAPE Algorithm in NMR Quantum Computer}



\author{Ebrahim Khaleghian$^{1}$, Arash Fath Lipaei$^{3}$, Abolfazl Bahrampour$^{1,2}$, Morteza Nikaeen$^{1,2}$, Alireza Bahrampour$^{1,2}$}

\affiliation{$^{1}$Department of Physics, Sharif University of Technology, Tehran 14588, Iran\\$^{2}$ Centre for Quantum Engineering and photonics Technology, Sharif University of Technology, Tehran, Iran\\$^{3}$ Computational Sciences and Engineering Department,  Koc University, Istanbul, Turkey}


\date{\today}

\begin{abstract}
In this paper, we introduce a neural network to generate optimal control pulses for general single-qubit quantum logic gates, within a Nuclear Magnetic Resonance (NMR) quantum computer. By utilizing a neural network, we can efficiently implement any single-qubit quantum logic gates within a reasonable time scale. The network is trained by control pulses generated by the GRAPE algorithm, all  starting  from  the  same  initial  point. After implementing the network, we tested it using numerical simulations. Also, we present the results of applying Neural Network-generated pulses to a three-qubit benchtop NMR system and compare them with simulation outcomes. These numerical and experimental results showcase the precision of the Neural Network-generated pulses in executing the desired dynamics. Ultimately, by developing the neural network using the GRAPE algorithm, we discover the function that maps any single-qubit gate to its corresponding pulse shape. This model enables the real-time generation of arbitrary single-qubit pulses. When combined with the GRAPE-generated pulse for the CNOT gate, it creates a comprehensive and effective set of universal gates. This set can efficiently implement any algorithm in noisy intermediate-scale quantum computers (NISQ era), thereby enhancing the capabilities of quantum optimal control in this domain. Additionally, this approach can be extended to other quantum computer platforms with similar Hamiltonians.
\end{abstract}

\maketitle

\section{Introduction}
Developing the software components of quantum computers plays a crucial role in achieving the ambitious goal of building scalable, fault-tolerant, and universal quantum computers. Today, alongside efforts to enhance the quality of the quantum layer and address the challenges of hardware architecture across various platforms, intense research focuses on quantum error correction, error mitigation, and quantum control methods. These methods help compensate for the unavoidable hardware defects.

One of the primary challenges in scaling up contemporary quantum computers is achieving high gate fidelities, which increases the depth of quantum circuits. Additionally, to achieve a fault-tolerant regime, gate fidelity must surpass a specific threshold. Enhancing gate fidelity beyond this threshold reduces the overhead qubits required for effective quantum error correction.

Several factors limit the fidelity of quantum logic gates, including the imprecision of classical control instruments, manufacturing defects in the quantum hardware layer, and qubit interactions with the environment, such as material imperfections. Additionally, some challenges arise from unwanted interactions within the quantum layer itself, even in closed quantum systems. For example, qubit cross-talk and the inevitable evolution of qubits due to qubit-mediated interactions during single-qubit operations present significant obstacles. These issues are prevalent across various platforms, including NMR and superconducting systems. This is where optimal quantum control methods can play a crucial role in enhancing gate fidelities. 

In multi-qubit quantum computing systems, achieving high-fidelity gate operations poses a significant quantum control challenge. In our work, this involves employing pulse shaping techniques to precisely guide the target qubit through its intended evolution while ensuring that non-target qubits remain effectively undisturbed.
However, inherent system couplings often induce unwanted interactions, leading to erroneous evolution of the target qubit and spurious dynamics in non-target qubits. To mitigate these effects, we engineer control pulses that direct the target qubit through its intended evolution while simultaneously steering non-target qubits along compensatory trajectories, such that their quantum states are restored by the conclusion of the target qubit’s operation. This strategy maintains the locality of the transformation, suppresses cross-talk, and enhances the overall fidelity of the control scheme by confining the unitary evolution to the intended subsystem.

The NMR platform is an excellent choice for testing quantum control ideas. First, bench-top NMR spectrometers are widely accessible in university laboratories, making them convenient for research. Moreover, due to their weak magnetic field, advanced pulse shaping is crucial for achieving optimal control in bench-top NMR quantum computers. Additionally, the results obtained can be readily applied and extended to other platforms with similar Hamiltonians, such as the control of superconducting qubits in a quantum bus resonator.

Quantum logic gates are implemented on the NMR platform by applying RF magnetic field pulses, with their central frequencies positioned close to the Larmor frequencies of different nuclei \cite{a1, a, b}. Extensive research has been conducted to determine the optimal pulse shapes required for high-fidelity quantum operations \cite{c, d, q, y, cc}. One approach to finding these pulse shapes for specific logic gates is optimal control theory, which offers several methodologies \cite{e, f, aa, bb, dd}. A particularly notable numerical technique within this framework is the GRAPE algorithm \cite{g}. This method is designed to optimize pulse shapes for basic quantum logic gates, such as S (phase gate), T (${\pi}/{8}$ gate), H (Hadamard gate), and the Pauli gates X, Y, and Z \cite{h}. Once the pulse shapes for these fundamental gates are generated—a process that is time-intensive—they are registered in the compiler. These registered pulses are then used to efficiently implement arbitrary gates within the algorithm.\\
A more sophisticated goal, particularly crucial in the NISQ era, is to extend this approach to designing pulse shapes for directly implementing arbitrary gates without decomposing them into basic ones—essentially enabling the execution of any arbitrary single-qubit gate with a single pulse \cite{i}. This is where neural networks come into play. By integrating the network into the compiler, pulse generation for arbitrary gates can be achieved within a reasonable time, i.e., during compile time.

In recent years, there has been a surge in research exploring the application of machine learning algorithms within the realm of quantum optimal control theory \cite{k, l, m, n, o, p, r, t, u, v, w, x, z}. Also, there has been significant research in developing NMR or MRI RF control pulse designs using neural networks by utilizing the excitation profile as a training set \cite{ii, jj, kk, ll, mm, nn}. Furthermore, neural networks have been explored for their role in the shimming of NMR spectrometers \cite{oo, pp, qq}, a feedback control process that ensures a homogeneous magnetic field across the NMR sample via various magnet coils. This process allows identical nuclei to experience uniform Zeeman splitting and have identical Larmor frequencies, resulting in narrower spectral lines. \\
In this work, we develop a neural network by employing the GRAPE algorithm for sampling, enabling the training of networks to implement logical gates in NMR quantum computers. The concept of control pulse generation can be further expanded through neural networks, as discussed in \cite{i, j}. Building on the foundational ideas presented in \cite{i, j}, we enhance the sampling method for training the neural network and test its effectiveness in an NMR quantum computing implementation. By leveraging a neural network, we can input the desired operation and obtain the corresponding pulse shape as output. This approach provides a versatile tool capable of generating pulse shapes for arbitrary gates in a reasonable time scale, overcoming the limitation of the GRAPE algorithm, which is constrained to optimizing pulse shapes for individual specific gates \cite{i, j}.\

To design such a network, a well-structured training set is required. For efficient training, we focus on generating phase-only control pulses \cite{rr, ss}. A straightforward strategy involves generating a series of random pulses and obtaining their resulting operations on the system. This process yields numerous ordered pairs, where the first element represents the operation and the second corresponds to the associated pulse shape. Using this training set, we can proceed to train a neural network.\

However, when we applied this approach to a three-qubit NMR system, the results were sub-optimal, with error rates failing to decrease satisfactorily. To address this issue, we explored utilizing the GRAPE algorithm for constructing a more effective training set. In this approach, we randomly selected a variety of single-qubit gates and employed the GRAPE algorithm to determine optimal control pulses for their implementation. A crucial aspect of our approach in using the GRAPE algorithm to train the network and achieve optimal results was ensuring that the optimization search started from a common initial point.\
Upon calculating the cosine similarity between the randomly generated pulses in our initial approach, we found that the similarity was relatively low. By instead using a common starting point in the GRAPE algorithm to develop pulses for the desired gates, we effectively constrained the search to a specific region within the space of control parameters. Consequently, within this region, the pulses exhibit a higher degree of similarity, as illustrated in Fig. 1.

\begin{figure}[htbp]
	\centering
	\includegraphics[scale=0.45]{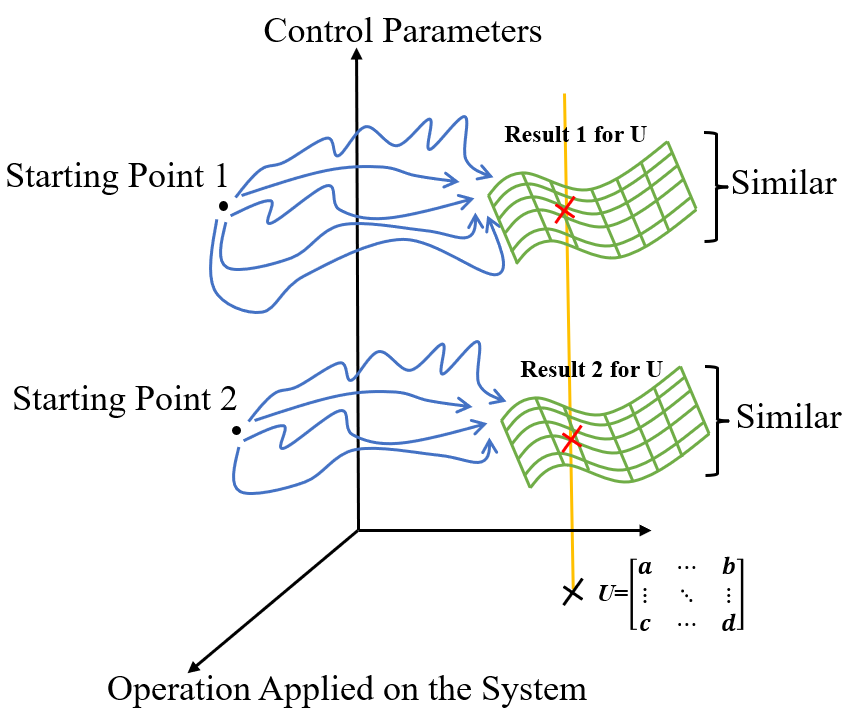}
	\caption{The subspace of the GRAPE-generated pulse shapes when the algorithm starts from the same point. In each subspace, pulses are similar to one another and correspond to a common starting point.}
\end{figure}

As we will demonstrate, employing this strategy significantly improves the cosine similarity between the resulting pulses. This approach helps us to focus on data around a specific point in the solution space. Intuitively, initializing the GRAPE algorithm from a random starting point for each operation leads to convergence toward very different pulse shapes. In contrast, using the same initial point for all operations results in similar pulse shapes across them. In a problem with multiple, or even infinitely many, local minima, it is crucial to limit our analysis of the dynamic system to a specific local solution. Ignoring this necessity results in severely non-smooth data geometry, preventing a simple, feasible neural network from finding any patterns within the feature space.\\
Therefore, initializing the GRAPE algorithm at a specific point before generating the training set is mandatory; otherwise, our mean squared error (MSE) regression optimization algorithm will struggle to fit the neural network to the data. Cosine similarity served as an appropriate criterion for analyzing the smoothness of our generated data. Initializing the GRAPE algorithm from the same point significantly increased this criterion, validating our strategy for generating a smooth training data set. Consequently, the development of a neural network with this training set proved successful.\\
Therefore, we have developed a neural network that, upon receiving a desired single-qubit gate as input, generates the corresponding pulse shape. Another key advantage of this approach is its ability to execute complicated operations within a single pulse. This method offers a significant improvement over sequential gate execution, particularly in the NISQ era, where accumulated gate fidelities can lead to a reduction in overall fidelity. By directly synthesizing the operation, we achieve higher fidelity with fewer steps.
\section{Hamiltonian of the system and form of the control pulses}
For an $n$-qubit NMR system with a homo-nuclear molecule, one can demonstrate that if the system is in the liquid regime at room temperature and operates with only one RF channel, then by selecting a rotating frame that matches the RF frequency of the channel, the total Hamiltonian in terms of the free evolution term and control term  can be expressed as follow \cite{ee, a, b, c}:

\setlength{\abovedisplayskip}{1pt}
\setlength{\belowdisplayskip}{1pt}

\begin{equation}
H_{\text{total}} = H_{\text{free}} + H_{\text{ctrl}},
\end{equation}

\begin{equation}
H_{\text{free}} = \sum_{i=1}^{n}(\omega_{0,i}-\omega_{\text{rf}}) \sigma_z^{(i)}+ \sum_{i<j} J_{ij}\sigma_z^{(i)}\sigma_z^{(j)},
\end{equation}

\begin{equation}
H_{\text{ctrl}} = \sum_{i=1}^{n} \omega_1(t) \left( \sigma_x^{(i)} \cos \phi + \sigma_y^{(i)} \sin \phi \right).
\end{equation}
Here, $\omega_{0,i}$ represents the Larmor frequency of qubit $i$ under the static magnetic field, while $\omega_{1}(t)$ denotes the time-dependent Larmor frequency of the qubits under the applied RF field, both arising from the Zeeman interaction and given by:
\begin{equation}
\omega_{0,i}=\gamma_{i} B_0, \omega_{1}(t)=\gamma_{i} B_1(t),
\end{equation}
where $B_0$ is the static magnetic field, and $B_1(t)$ and $\phi(t)$ represent the amplitude and phase of the applied RF field, respectively. $\gamma_{i}$ is the gyromagnetic ratio for each nucleus in the system. Since $B_1(t)$ is significantly smaller than $B_0$, and the system is homonuclear, we use the approximation $\gamma_{i} B_1(t) = \omega_{1}(t)$ for any $i$. The second term in $H_{\text{free}}$ represents the J-coupling between different nuclei \cite{b}.

To implement the desired quantum logic gates, we must determine \( B_1(t) \) and \( \phi(t) \). As a common technique in the GRAPE algorithm, these pulses are approximated as constant within $N$ discrete time steps. Within each step, the amplitude and phase remain constant, allowing us to compute the system's overall operation by multiplying the operations from each time step. Since the Hamiltonian is constant within each step, we can readily calculate the dynamics for that period. Consequently, the total operation resulting from the pulse sequence is obtained by the product of the dynamics across all steps \cite{ff}, 

\begin{equation} \label{eq5}
U_{total}=u_1u_2.....u_{N-1}u_N
\end{equation}
where \( u_i \) represents the unitary operation applied in step \( i \). In this algorithm, the derivative of fidelity with respect to control parameters is computed. By iteratively updating these parameters and following the gradient direction, we can converge on the optimal solution \cite{h}. Here, we focus solely on implementing a single-qubit gate. The objective is for the overall system dynamics to produce the desired operation on the target qubit, while keeping all other qubits unaffected—i.e., they should undergo no evolution, that is:
\begin{equation} \label{eq6}
U_{target}=U_1  \otimes I_2  \otimes I_3,
\end{equation}
where \( U_1\) represents the effective evolution of qubit 1 after \( N \) steps, and \( I_j \) describes the effective evolution of qubit \( j \), ensuring that its state returns to the initial state—i.e., no evolution occurs for that qubit. Our demand is that $U_{total}$ approaches the $U_{target}$ as closely as possible.

To simplify the learning process and reduce the number of control parameters, we used only phases to develop control pulses while keeping the amplitude constant. As we will see (and discussed in \cite{rr, ss}), this constraint does not significantly limit us in achieving acceptable fidelities, and we can reach high fidelities by only modulating the phase. To test the stability of this approach, after calculating the phase-modulated pulse with constant amplitude, we used it as the initial state for the GRAPE algorithm, by adding the amplitudes as an additional control parameters. The final result was not significantly different from the initial state, indicating that these pulses approximately satisfy the condition of zero gradient Euclidean norm with respect to phase elements, even after additional degrees of freedom related to amplitudes are added to the optimization problem.
\section{Designing the neural network}

As previously discussed, efficiently training the target neural network requires a set of sample pulses with high cosine similarity, ensuring that the pulses remain closely related. Cosine similarity quantifies the resemblance between two vectors by measuring the cosine of the angle between them in a multi-dimensional space. A value closer to 1 signifies a higher degree of similarity, indicating that the vectors—here, the control pulses—are structurally more alike. 
For efficient training, we focus on generating phase-only control pulses \cite{rr, ss}. In this approach, each pulse is represented as a vector, with its elements corresponding to the phase at specific time intervals. In this way, cosine similarity serves as a key metric for evaluating the similarity between sample pulses, enabling precise assessment and aiding in pulse selection for network training.

A straightforward strategy for training the network involves generating a series of random pulses and computing the resulting operations they induce in the system. However, upon applying Algorithm 1 to compute the cosine similarity between these generated pulses, we observed that their similarity was not particularly significant. This highlights the need for a more refined approach to pulse selection to enhance network performance.

\begin{algorithm} [H]
\caption{Cosine Similarity Analysis for the Control Pulses}
\begin{algorithmic}[1]
\For{$i \gets 1$ to $N_{\text{total}}$}
    \For{$j \gets 1$ to $N_{\text{total}}$}
        \State \( c[i,j] \gets \dfrac{\langle \phi_{\text{data}}[i],\, \phi_{\text{data}}[j]\rangle}{\|\phi_{\text{data}}[i]\|\cdot \|\phi_{\text{data}}[j]\|} \)
        \Comment{Compute the cosine similarity between pulse sequences of sample \(i\) and \(j\).}
    \EndFor
\EndFor
\end{algorithmic}
\end{algorithm}\

We then consider utilizing the GRAPE algorithm to develop a training set for the neural network. Hereafter, we restrict ourselves to a three-qubit system and train the network to generate pulses for implementing any desired gate on qubit 1. Training the network to generate pulses for arbitrary gates on other qubits (i.e., qubits 2 and 3) follows the same procedure.

\begin{algorithm}[H]
\caption{Input Data Generation Using Axis–Angle Representation with Cosine–Inverse Sampling}
\begin{algorithmic}[1]
\For{$i \leftarrow 1$ to $N_{\text{total}}$}
    \State Generate a uniformly random $2\times2$ unitary matrix $U_1 \in SU(2)$
    \State Sample random variables:
    \[
    \phi \sim \mathcal{U}(0, 2\pi), \quad
    r \sim \mathcal{U}(0, 1), \quad
    \gamma \sim \mathcal{U}(0, 2\pi)
    \]
    \State Compute the polar angle using the cosine–inverse approach:
    \[
    \theta = \arccos(1 - 2r)
    \]
    \State Define the unit vector representing the rotation axis on the Bloch sphere:
    \[
    \hat{n}(\theta,\phi) =
    \begin{pmatrix}
    \sin\theta \cos\phi \\
    \sin\theta \sin\phi \\
    \cos\theta
    \end{pmatrix}
    \]
    \State Construct the unitary matrix using the axis–angle representation:
    \[
    U_1 =  \cos\!\left(\frac{\gamma}{2}\right) I_2
      - i \sin\!\left(\frac{\gamma}{2}\right)
      \left(n_x \sigma_x + n_y \sigma_y + n_z \sigma_z\right)
    \]
    \State Embed the $2\times2$ unitary into an $8\times8$ unitary matrix via Kronecker products
    \[
    U \leftarrow \mathrm{kron}(\mathrm{kron}(U_1, I_2), I_2)
    \]
    \State Store real and imaginary parts
    \[
    X[i,:,:,1] \leftarrow \Re(U), \quad X[i,:,:,2] \leftarrow \Im(U)
    \]
\EndFor
\end{algorithmic}
\end{algorithm}

In first scenario, we used the GRAPE algorithm to obtain the optimal control pulses for implementing a typical single-qubit gate (Hadamard gate), each initialized from a different starting point.  Figure 2 illustrates the cosine similarity of 16 samples generated using the GRAPE algorithm for a specific logic gate, each initialized with a different starting point. The similarity among pulses generated by the GRAPE algorithm for the same gate but with different starting points is relatively low. 

In the second scenario, we selected 17,000 uniformly chosen single-qubit gates (Algorithm 2) and applied the GRAPE algorithm to determine their optimal control pulses, maintaining a common starting point across all instances. To construct this uniform set, we adopt the axis-angle representation of \( U(2 \times 2) \) unitary matrices in the Bloch sphere, defined as  $U(\theta, \phi, \gamma) = e^{-i \sigma \cdot \hat{n}(\theta, \phi) \gamma/2} $, where \( \hat{n}(\theta, \phi) \) represents the rotation axis and \( \gamma \) denotes the rotation angle in the Bloch sphere representation. By uniformly selecting $\hat{n}(\theta, \phi)$ on the sphere using the cosine inverse approach and also uniformly selecting the rotation angle $\gamma$ around the axis we construct the desired set of uniform unitary matrices. Surprisingly, as shown in Fig. 3, the cosine similarity analysis of these 17,000 samples reveals a high degree of similarity among the pulses, despite their correspondence to different gates. A comparison between Figures 2 and 3 reveals a noticeable increase in similarity among the samples when the GRAPE algorithm is used with a common starting point. This highlights the importance of maintaining a fixed initialization when generating training pulses, as it effectively constrains the search to a specific region within the space of control parameters, ensuring that the resulting pulses remain closely related.\\

\begin{figure}[htbp]
	\centering
	\includegraphics[width=\linewidth]{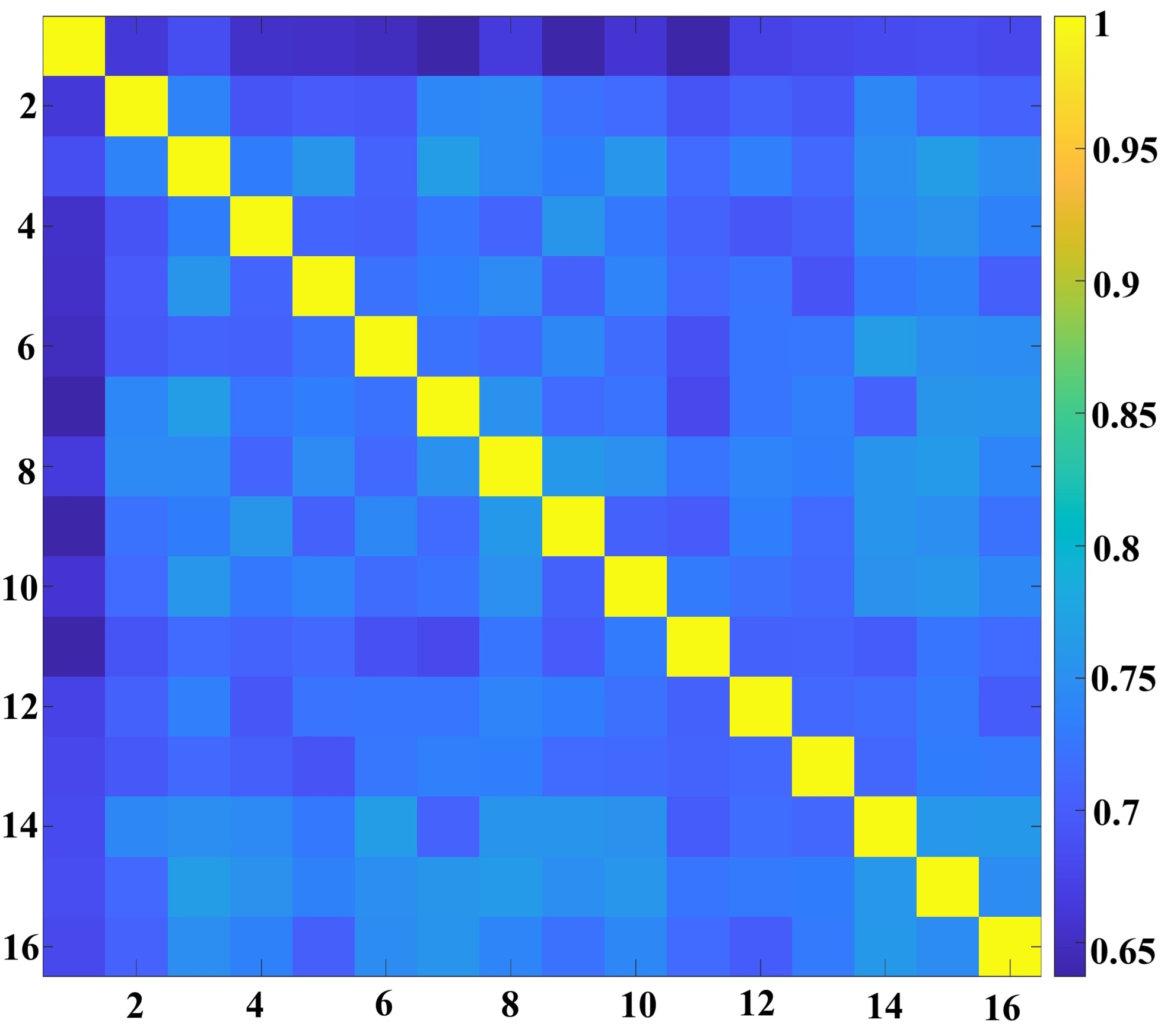}
	\caption{Cosine similarity analysis of 16 samples generated using the GRAPE algorithm for a specific logic gate, each with a different starting point.The horizontal (x) and vertical (y) axes represent the indices of sample pulses. Value of the similarity between the pulses x and y is shown on the x-y cells. The diagonal entries all equal 1, as the cosine similarity between a pulse and itself is always 1.}
\end{figure}
\begin{figure}[htbp]
	\centering
	\includegraphics[width=\linewidth]{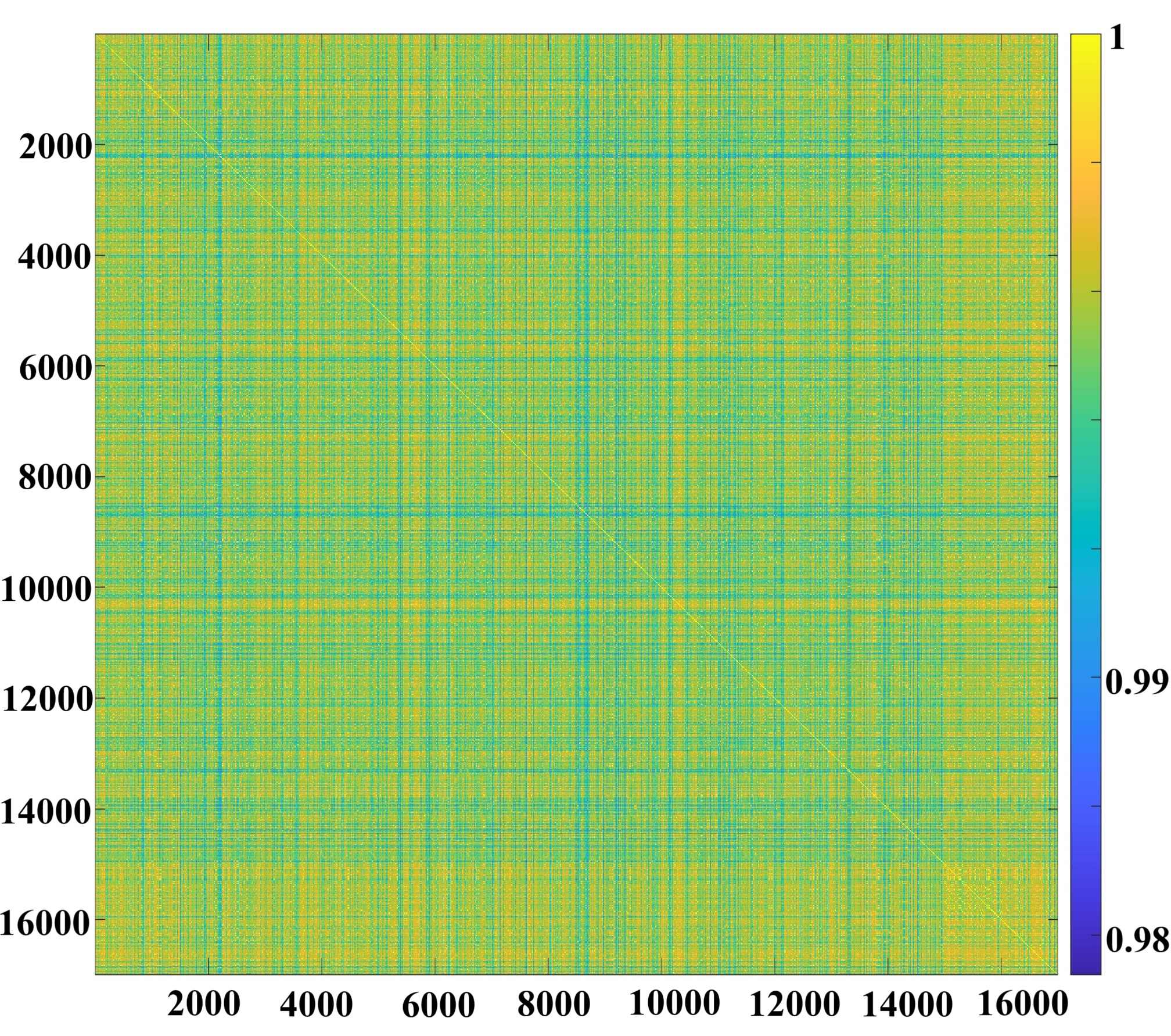}
	\caption{Cosine similarity analysis of 17,000 samples generated using the GRAPE algorithm for the random logic gates, each with a same starting point.The horizontal (x) and vertical (y) axes represent the indices of sample pulses. Value of the similarity between the pulses x and y is shown on the x-y cells.}
\end{figure}

Therefore, we adopted the second scenario to generate sample pulses corresponding to selected unitary operations, forming the ground truth for training data-set, ensuring efficient network training (Algorithm 3).

\begin{algorithm}[H]
\caption{Output Ground Truth for Training Set}
\begin{algorithmic}[1]
\ForAll{Unitary \(i\) \textbf{in X}}
    \State \(G_{\text{target}} \gets\)\(X[i,:,:,1]+j.X[i,:,:,2]\), where j is the imaginary unit.
    \State Initialize control parameters: \(A \gets A_0\) and \(\phi \gets \phi_0 = \begin{bmatrix} \pi & \pi & \pi & \dots & \pi \end{bmatrix}\)
, where \(A_0\) is assumed to be uniform on time.
    \For{\(iter \gets 1\) to \(num\_iterations\)}
        \State \(d\phi \gets \textsc{Deri\_func}(A_0, \phi, G_{\text{target}})\)
        \Comment{Compute gradients of the fidelity cost with respect to \(\phi\), using the methodology in \cite{c}. Treating \(A_0\) not to be a part of the optimization set.}
        \State Update \(\phi\) using  \(d\phi\) and ADAM Gradient Ascent \cite{ttt}.
    \EndFor
    \State Save optimized pulse parameters: \(\phi_{\text{data}}[i] \gets \phi\).
\EndFor
\end{algorithmic}
\end{algorithm}

After creating 17,000 samples, we trained a feed-forward neural network \cite{Gdf} to establish the mapping function between the overall operation applied to the system and control pulses (Algorithm 4). The trained neural network is shown in Fig. 4. To implement the layers of the neural network, we utilized the architecture depicted in Fig. 5.
\begin{algorithm}[H]
\caption{Neural Network Design and Training for Pulse Generation}
\begin{algorithmic}[1]
\State Inp $\gets$ For each generated gate, U (the 2×2 matrix), flatten and concatenated the real and imaginary components, into a one-dimensional input array with 8 elements.
\State Create the complete dataset structure \((Inp, \phi_{\text{data}})\), and split the dataset into training set and test set.
\State Define a deep feed-forward neural network with:
\begin{itemize}
    \item \textbf{Input layer:} 8 features.
    \item \textbf{Hidden layers:} Several fully-connected layers with non-linear activations (e.g., tanh, ReLU) and dropout layers.
    \item \textbf{Output layer:} \(T\) real-valued outputs corresponding to the predicted pulse phases, T is the length of the phase pulse, $\phi$.
\end{itemize}
\State Set hyperparameters and use back propagation to train the network. \Comment{Repeat by using various architectures and combinations of hyperparameters, until specific convergence criteria are met. Regulate the dropout rate and L2 norm to prevent overfitting.} \Comment{The network learns the mapping from unitary features to optimal pulse parameters.}
\end{algorithmic}
\end{algorithm}
\begin{figure}[htbp]
	\centering
	\includegraphics[width=\linewidth]{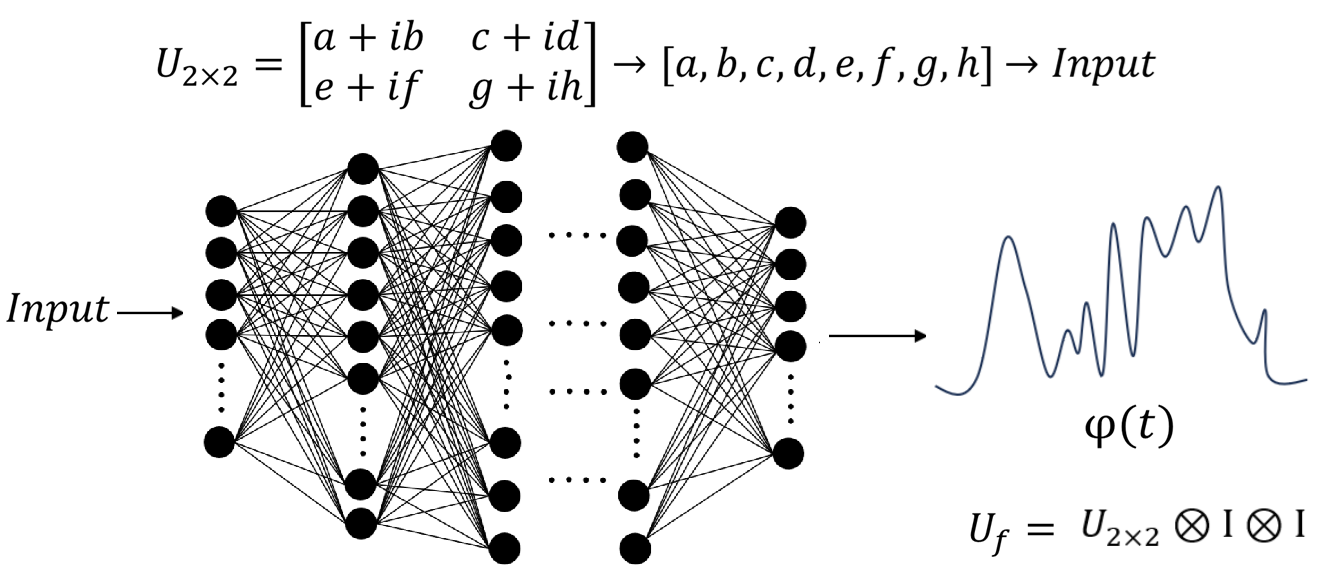}
	\caption{The neural network receives a \( U \) (2×2 unitary matrix), representing the gate to be applied to the target qubit (for example, qubit 1). It flattens and concatenates the real and imaginary components into a one-dimensional input array with 8 elements. The network then interprets the overall evolution as \( U_1 \otimes I_2 \otimes I_3 \) and generates the pulse required to implement the desired gate, which is transmitted to the output. Upon applying the pulse, the operator \( U_f \) is executed on the system.}
\end{figure}
\begin{figure}[htbp]
	\centering
	\includegraphics[width=\linewidth]{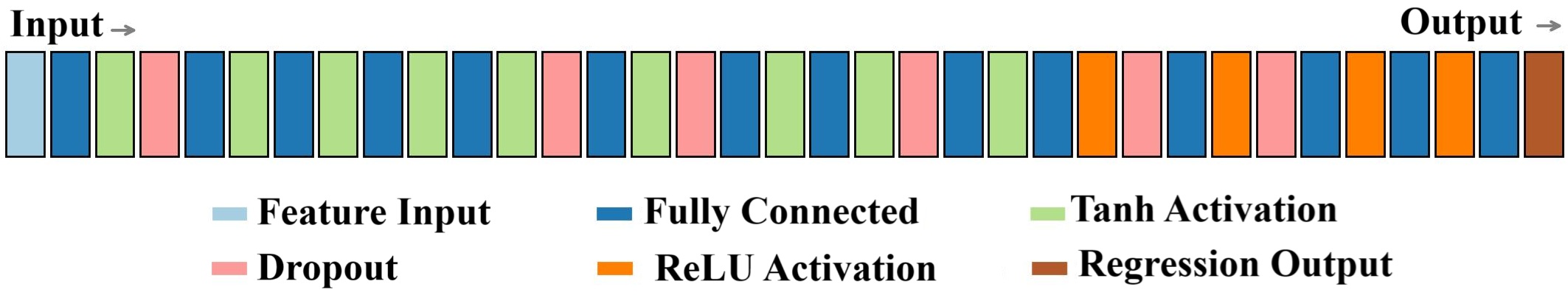}
	\caption{Neural Network Architecture Overview. This diagram presents the structured flow of the neural network, beginning with the feature input and advancing through sequential layers that include fully connected layers, dropout modules, and activation functions (ReLU and Tanh). The final stage culminates in a regression output, illustrating how the network processes and transforms input data to generate predictions.}
\end{figure}
\section{Testing the neural network}

To evaluate the performance of the developed neural network, we conduct a series of numerical experiments. For the primary assessment, the network is tested on a dataset of 15,000 uniformly distributed \( 2 \times 2 \) unitary matrices generated using Algorithm 2.

Finally,  we  feed  these  unitary  matrices  into  the trained  network  and  compute  the  operation  that would  be  applied  to  the  system  upon  implementation of the neural network-generated pulse. To determine the total resultant operation, we employ the same approach used in the GRAPE algorithm \cite{ff}, as detailed in Eqs.\ref{eq5} and \ref{eq6}.
The fidelity was then calculated between $U_{Target}$, the unitary operation intended as the input to the neural network, and $V$, the unitary operation actually implemented by the pulse generated by the neural network, using the definition:
\begin{equation}
F(U_{Target},V)=Tr(U_{Target}^\dagger V).
\end{equation}
Next, we plot the histogram of the fidelity, as illustrated in Fig.6, which demonstrates the perfect quality of the results.
\begin{figure}[htbp]
	\centering
	\includegraphics[width=\linewidth]{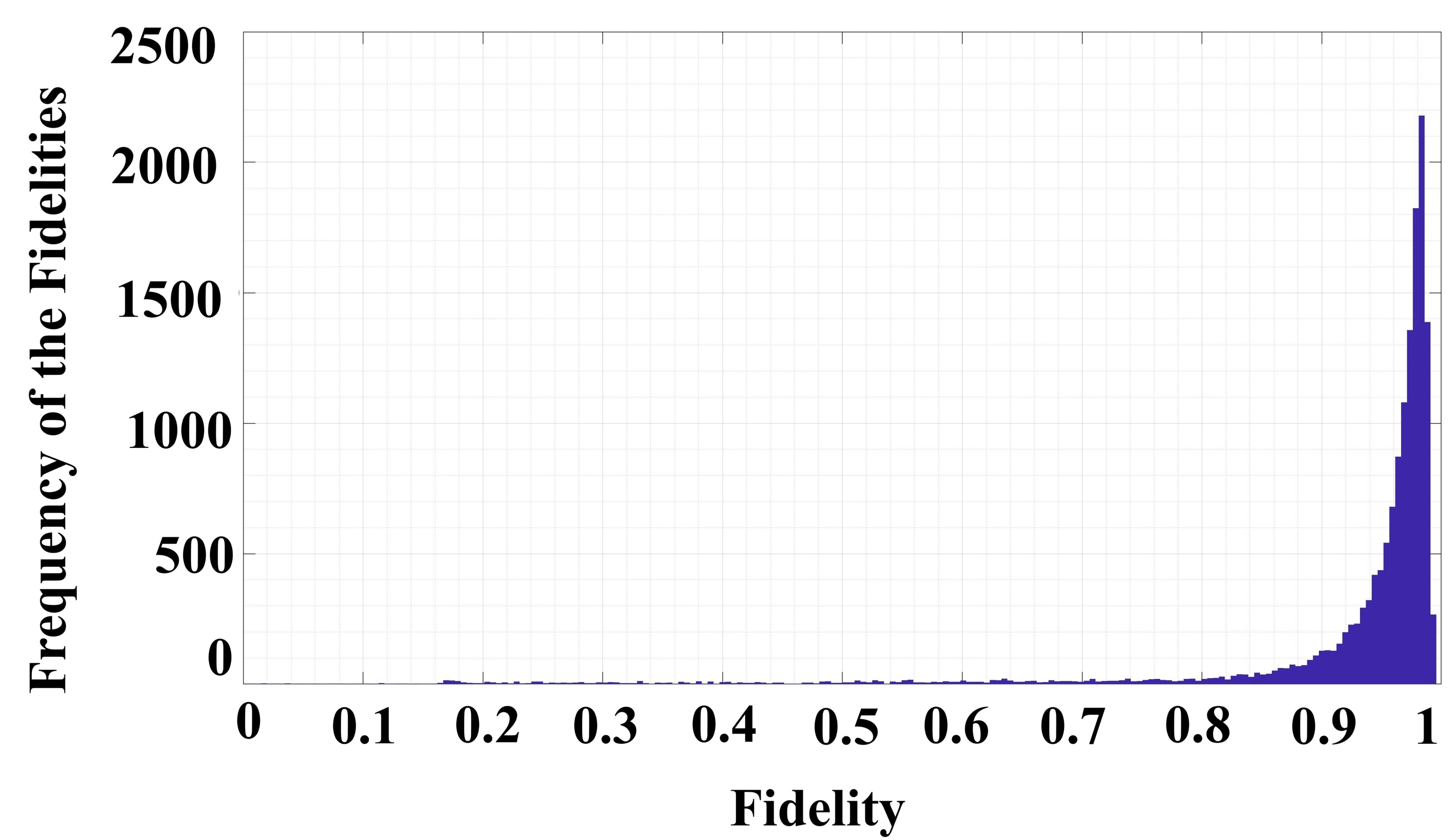}
	\caption{Histogram illustrating the distribution of fidelities in the test results obtained from the neural network's output. The data-set comprises 15,000 samples, with the majority of gates achieving a fidelity greater than 0.9.}
\end{figure}
\\
In Table.1, you'll find information on the generated pulses. As illustrated, this method is significantly faster than the standard GRAPE approach, which typically takes at least 15 seconds. Consequently, this method is approximately three orders of magnitude faster, and can be integrated as part of the compiler in quantum computer architectures.
\begin{table}
\begin{tabularx}{0.48\textwidth} { 
  | >{\centering\arraybackslash}X 
  | >{\centering\arraybackslash}X 
  | >{\centering\arraybackslash}X | }
 \hline
 Mean of the Fidelities & Standard Deviation of the Fidelities & Access Time of the Neural-Network \\
 \hline
 92.5\%  & 13.2\%  & 14.8ms  \\
\hline
\end{tabularx}
\caption{Performance Quality of the Multi-Layer Perceptron.}
\end{table}

As a second illustration of the network's performance, we conducted tests on the following gates using the network, where $H$ represents the Hadamard gate:
\begin{equation} \label{eqhxyz}
\begin{split}
    U_x(\theta) &= H e^{-i\theta\sigma_x^{(1)}}, \\
    U_y(\theta) &= H e^{-i\theta\sigma_y^{(1)}}, \\
    U_z(\theta) &= H e^{-i\theta\sigma_z^{(1)}}.
\end{split}
\end{equation}

The gate fidelities corresponding to unitary transformations given in Eq.\ref{eqhxyz} are plotted against rotation angle $\theta$ in Fig.7.\\

\begin{figure}[htbp]
	\centering
	\includegraphics[width=\linewidth]{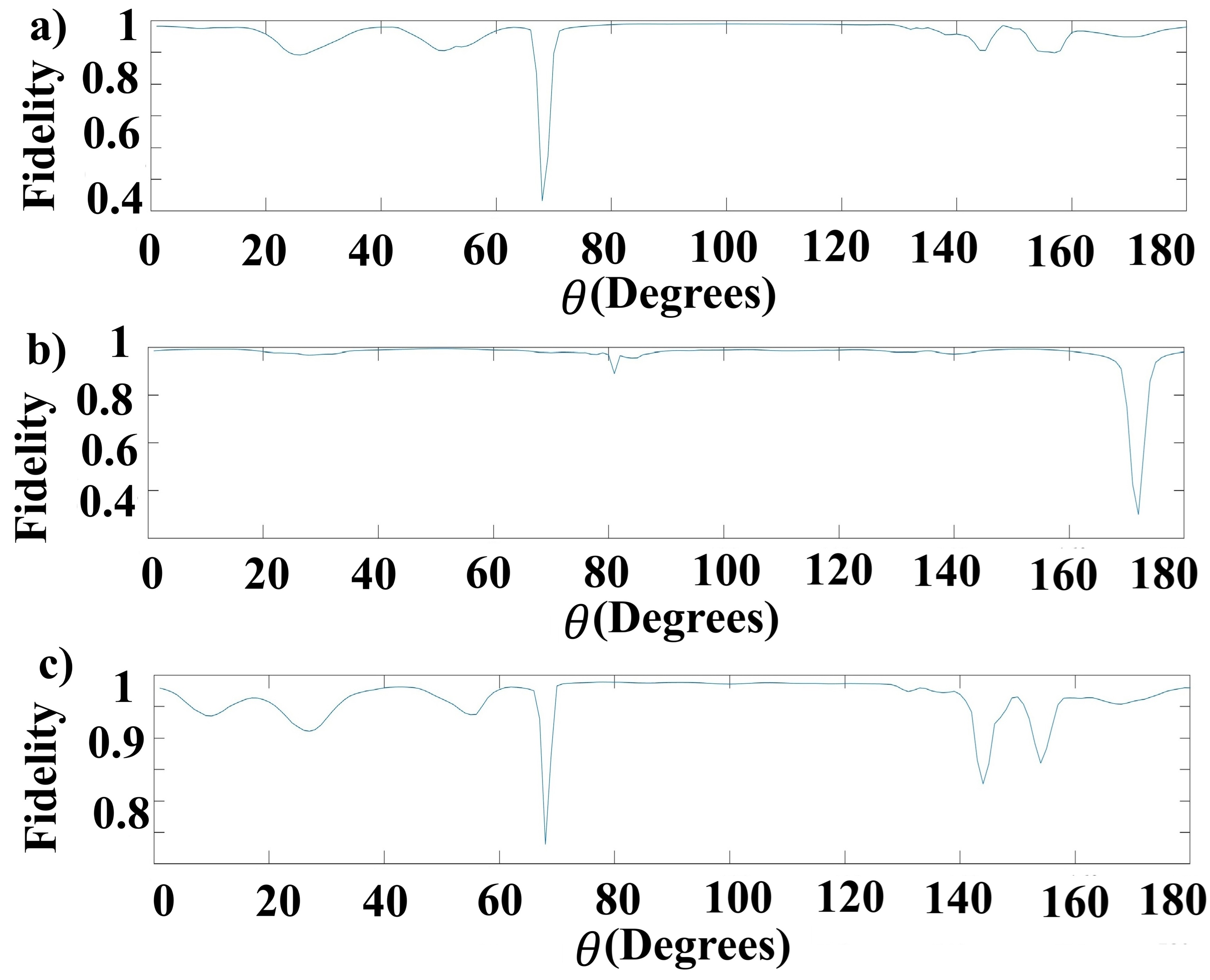}
	\caption{Fidelity of unitary transformations given in Eq. \ref{eqhxyz} as a function of \( \theta \), derived from the output of the trained network. Panels (a), (b), and (c) correspond to the operations \( U_x(\theta) \), \( U_y(\theta) \), and \( U_z(\theta) \), respectively.}
\end{figure}
Finally, we examine the fidelity results of the pulses generated by the network for unitary transformations corresponding to different axes and different angles as given by:
\begin{equation} \label{eqalphateta}
U(\theta,\alpha)=e^{-i\theta(\cos(\alpha)\sigma_x^{(1)}+\sin(\alpha)\sigma_y^{(1)})}.
\end{equation}
The gate fidelities corresponding to the unitary transformation given in Eq.\ref{eqalphateta} is plotted against $\theta$ and $\alpha$ in Fig.8 and Fig.9.
\begin{figure}[htbp]
	\centering
	\includegraphics[width=\linewidth]{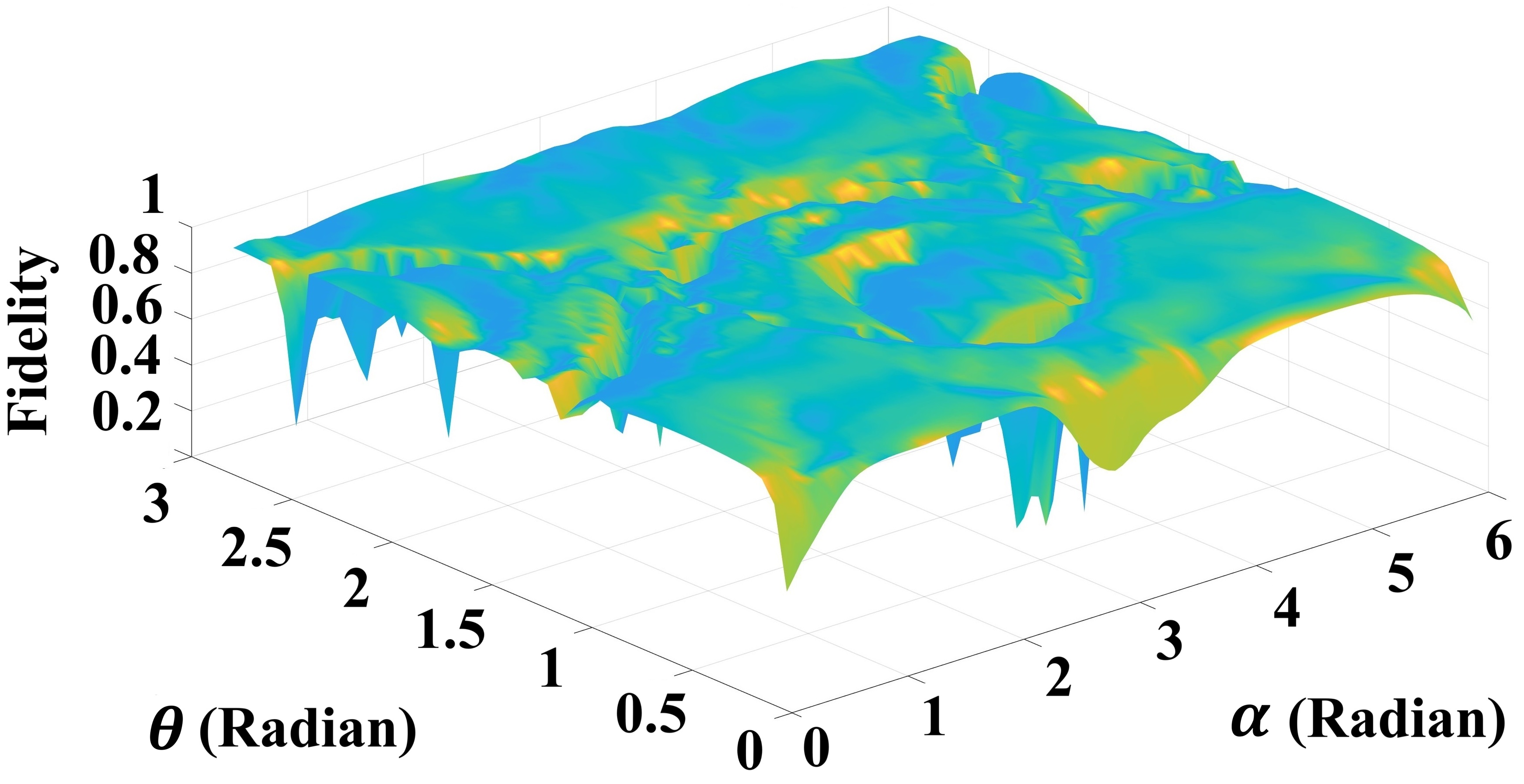}
	\caption{The gate fidelities corresponding to the unitary transformation given in Eq.\ref{eqalphateta} is plotted against $\theta$ and $\alpha$, for pulses obtained from the output of the trained network. $\theta$ ranges from $0^{+}$ to $\pi^{-}$, and $\alpha$ ranges from $0^{+}$ to 2$\pi^{-}$, with a step size of $\frac{\pi}{50}$ for both angels.}
\end{figure}
\begin{figure}[htbp]
	\centering
	\includegraphics[width=\linewidth]{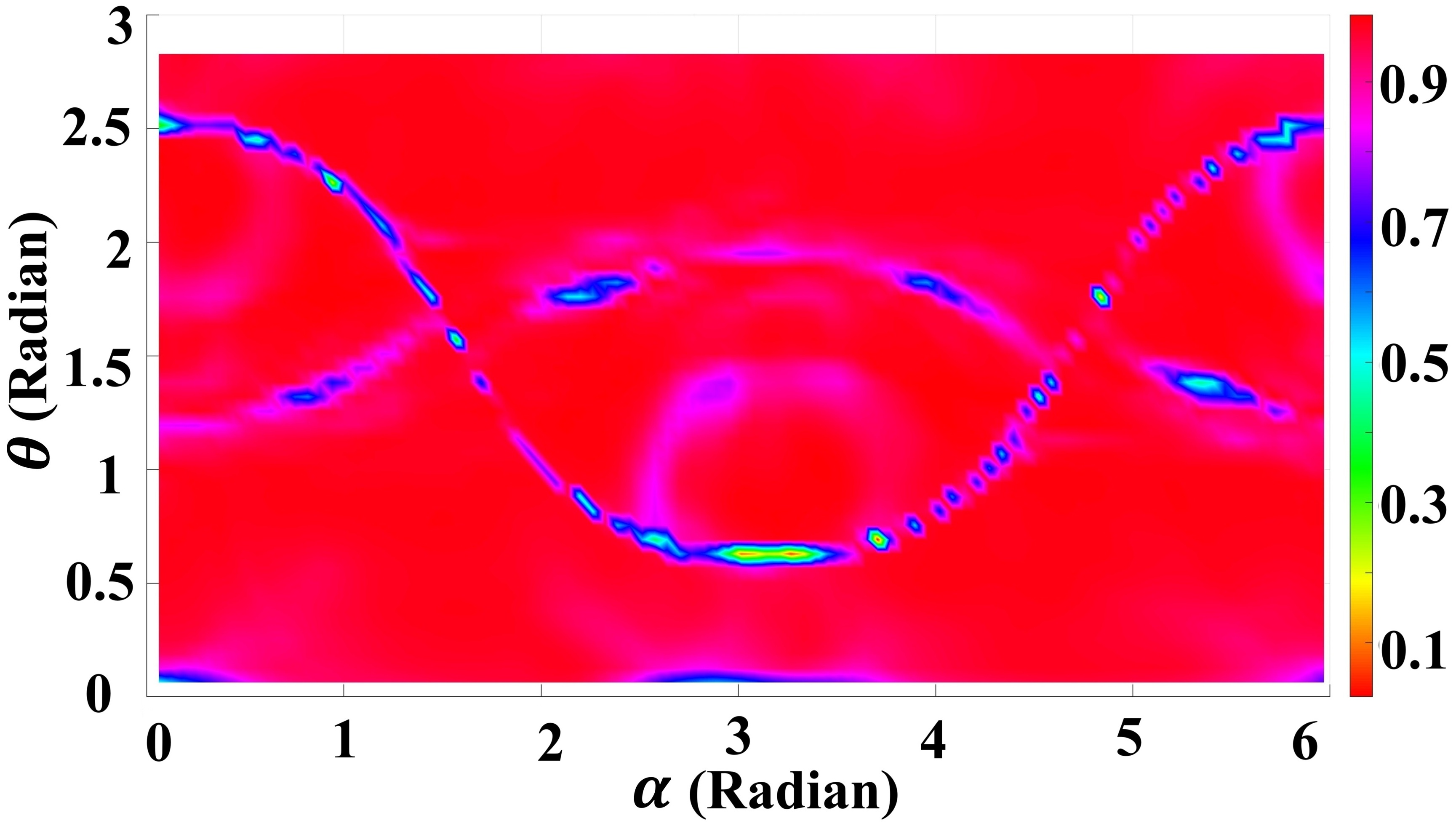}
	\caption{2D representation of the Fig.8}
\end{figure}

The mean and standard deviation of the fidelities for implementing the unitary transformations described in Eq. (6) using our network are $95\%$ and $8\%$, respectively. As illustrated in Figs. 7 and 8, noticeable dips in fidelity are observed. A possible explanation is that limited sampling coverage, combined with the inherent nonlinearity of the underlying physical system, may be more pronounced for many of these low-fidelity gates, thereby making the training process more challenging. Nonetheless, increasing the network depth did not lead to a significant improvement in fidelity, suggesting that the learning dynamics have saturated under the current approach.

\section{Experimental Test}
To ensure that the neural network will perform properly in real experiments, we experimentally implemented the generated pulses for quantum logic gates of Eq.\ref{eqalphateta} on a real benchtop NMR system.
 In NMR spectroscopy, the average value of the magnetization in the plan perpendicular to $B_0$, is measured and subsequently subjected to a Fourier transformation. It has been shown that within the frequency domain, the observed peaks correspond proportionally to specific elements of the system’s density matrix \cite{a, b}. This process is thoroughly discussed in quantum state tomography on an NMR systems \cite{gg}. Our NMR system is a $1$ $Tesla$ benchtop spectrometer with $C_2F_3I$ molecule, with fluorine nuclei serving as qubits. At $1$ $Tesla$, we observed the following constants of the experiment:
\begin{equation} \begin{aligned} \omega_1 - \omega_{rf} = -1375 \, \text{Hz}, \\ \omega_2 - \omega_{rf} = 56 \, \text{Hz}, \\ \omega_3 - \omega_{rf} = 1035 \, \text{Hz} \end{aligned} \end{equation} \begin{equation} \begin{aligned} \omega_{rf} = 40 \, \text{MHz}, \\ J_{1,2} = -67 \, \text{Hz}, \\ J_{1,3} = 28 \, \text{Hz}, \\ J_{2,3} = 38 \, \text{Hz}, \end{aligned} \end{equation}
\begin{figure}[htbp]
	\centering
	\includegraphics[width=\linewidth]{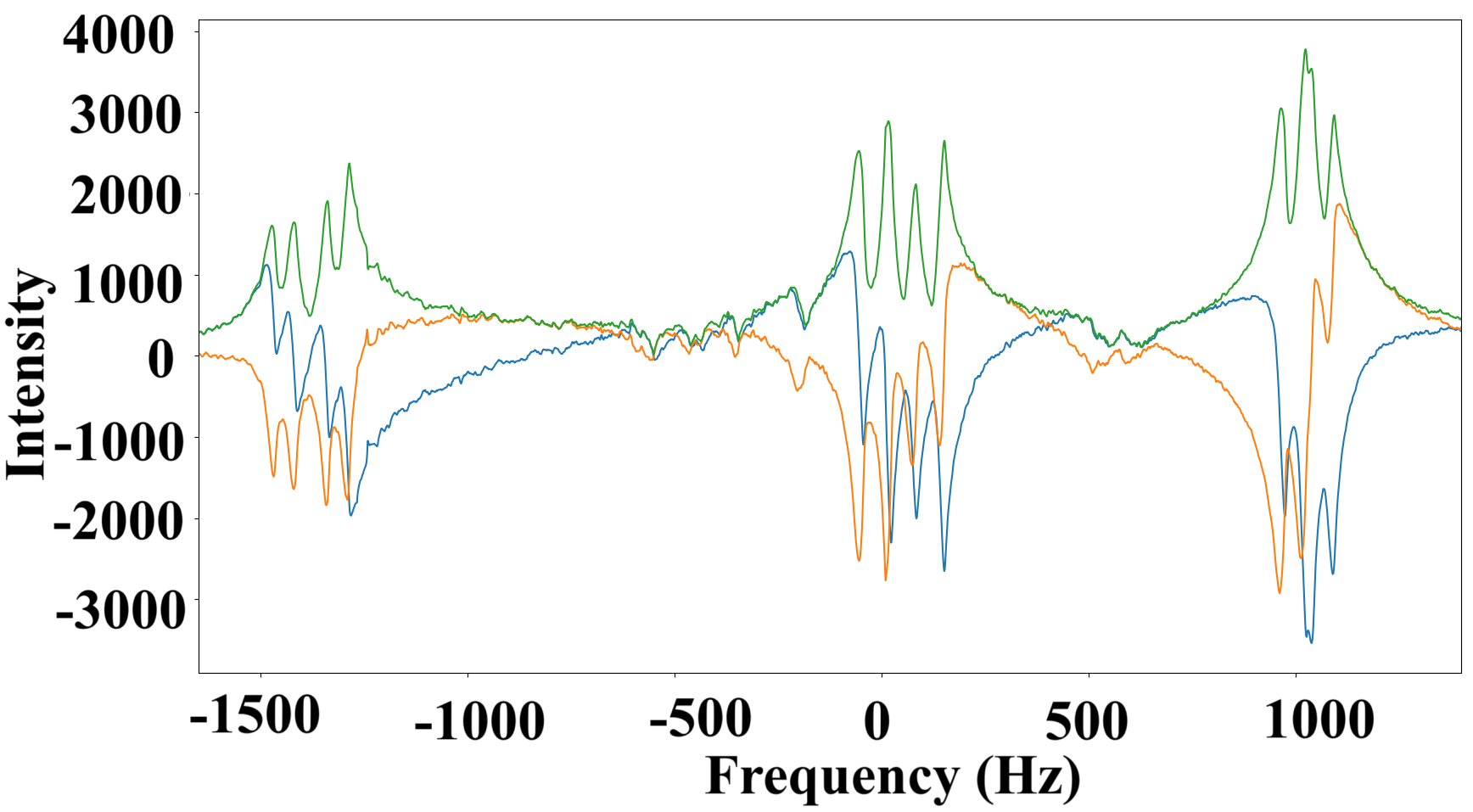}
	\caption{The relative intensity of the output spectrum of the NMR device is plotted against the frequency (in Hz), with the frequency axis shifted from 40 MHz. The blue curve represents the real part of the Fourier transform, the orange curve depicts the imaginary part, and the green curve indicates the total intensity.}
\end{figure}
\begin{figure}[htbp]
	\centering
	\includegraphics[width=\linewidth]{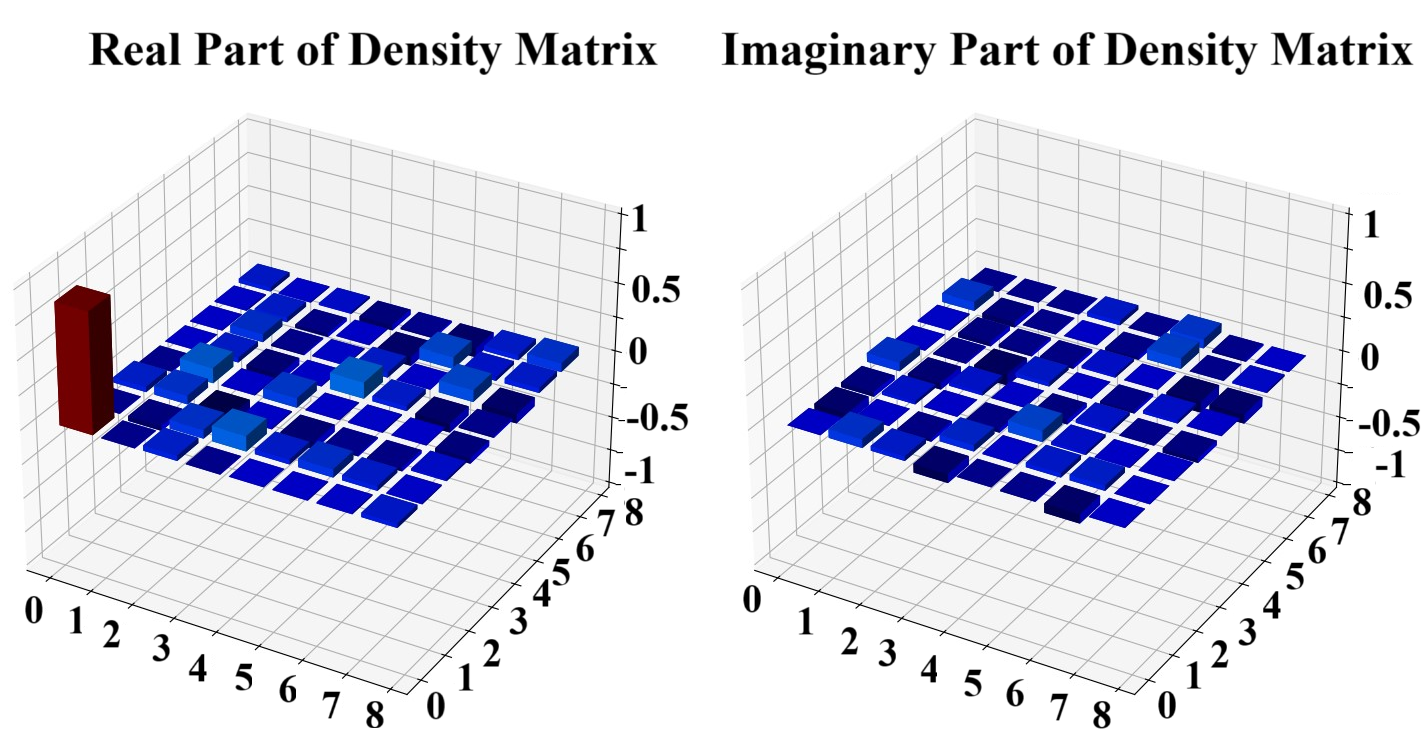}
	\caption{The tomography results, obtained after the preparation of the pseudo-pure state, with 0.89 fidelity}
\end{figure}
where $\omega_{i}$ represents the Larmor frequency of qubit $i$, $\omega _{rf}$ is the frequency of applied RF field and $J_{ij}$  represents the J-coupling coefficients between qubit $i$ and qubit $j$.  Fig.10 depicts the output spectrum of the system under applying the simple square shape pulse to the thermal state of the sample. For wide spectral width of spins one should check that square shaped RF pulse must able to homogeneously excite all the three spins. one example of uniform excitation of spins with wide spectral width can be found in \cite{dogra2015experimental}. In the context of NMR spectroscopy for a 3-qubit system with fully coupled nuclei, one would expect to observe 12 distinct peaks \cite{gg, a, b} which is clearly demonstrated in the output spectrum of Fig.10. 
To demonstrate the precision of the neural network, we generate pulses corresponding to gates of the form given in Eq.\ref{eqalphateta} with various values of $\alpha$ and $\theta = \frac{\pi}{4}$ and then consider the spectrum around qubit 1 after applying the generated pulses on the pseudo-pure state (PPS) \cite{uuuu}.

To prepare a three-qubit pseudo-pure state, we employed the relaxation method described in Ref. \cite{wwww, vvvv}. The process begins with the thermal equilibrium state, characterized by a Boltzmann distribution at room temperature and then leverages repeated applications of basis permutation operations combined with \( T_1 \) relaxation. The permutation gate, \( U_{\text{permute}} \) redistributes the populations among the basis states \( \ket{001}, \ket{010}, \ket{011}, \ket{100}, \ket{101}, \ket{110}, \) and \( \ket{111} \), while leaving \( \ket{000} \) unchanged. The permutation operation alternates with relaxation intervals, allowing \( T_1 \) relaxation to take effect. After multiple cycles, the system stabilizes, with the \( \ket{000} \) basis state being predominantly occupied and the remaining basis states sharing a lower, equal probability.  The resulting state constitutes the PPS and serves as the initial \( \ket{000} \) state.
Additionally, it should be noted that we utilized GRAPE-generated pulses for both tomography and permutation gate in PPS generation. The tomography result of the generated PPS is illustrated in Fig. 11, providing a visual representation of the reconstructed quantum state.

It has been shown that the integrals of the real and imaginary components of the output spectrum near the resonance frequencies determine the density matrix elements of the system \cite{a, b}. Figure 12 presents the simulated real and imaginary parts of the NMR output spectrum for a pseudo-pure state associated with qubit one, under applied pulses corresponding to the operations defined in Eq.~\ref{eqalphateta}, with \( \theta = \frac{\pi}{4} \) and various values of \( \alpha \). Since the single-qubit gate is applied exclusively to qubit one, we focus on monitoring the spectral evolution near frequencies close to its Larmor frequency. Figure 13 shows the corresponding experimental results.

\begin{figure}[htbp] \label{simuresults}
	\centering
	\includegraphics[width=\linewidth]{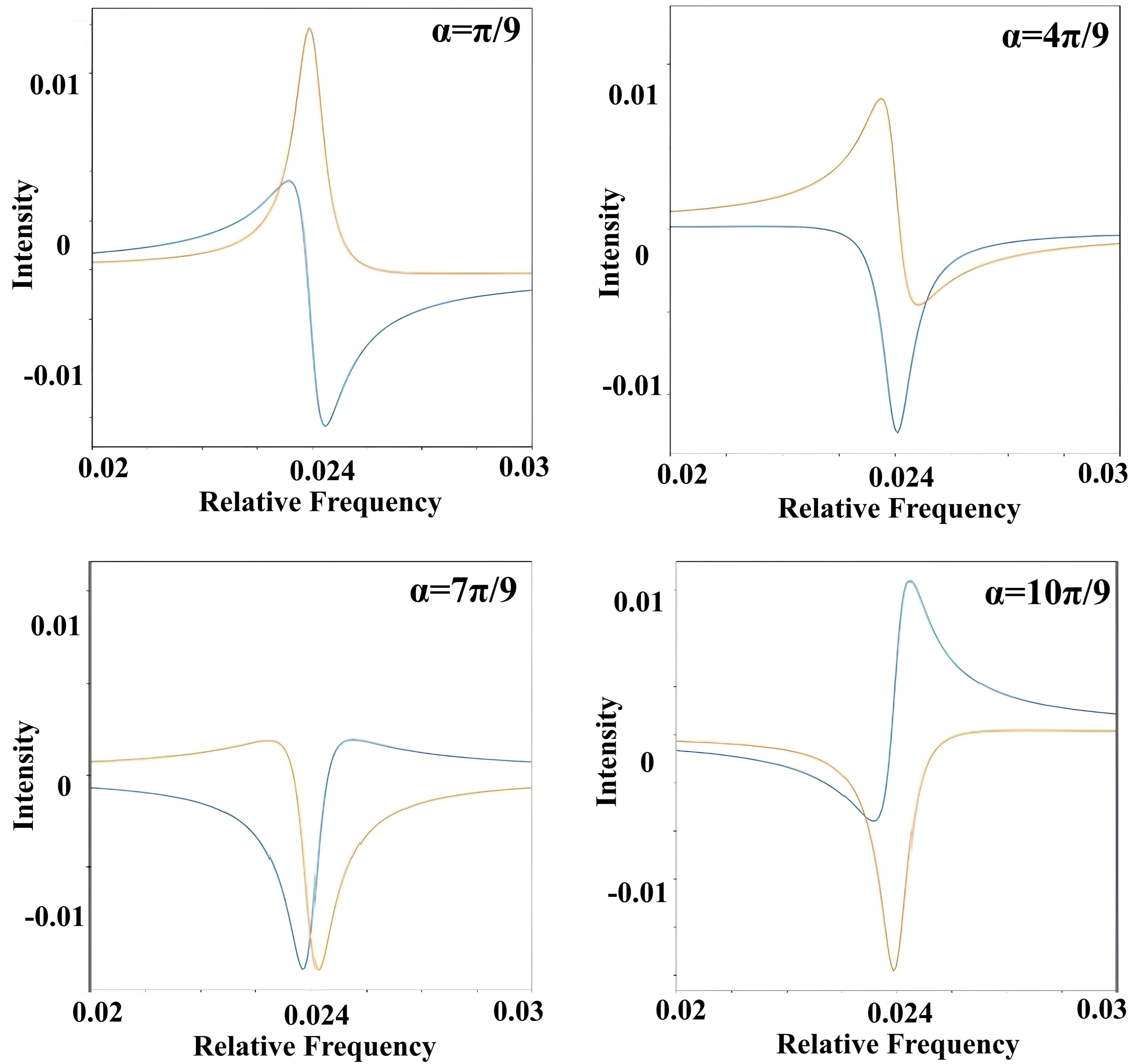}
	\caption{Simulation results of the NMR spectrum around qubit one after applying gates in Eq.\ref{eqalphateta} on pseudo-pure state for $\theta=\frac{\pi}{4}$ and different values of $\alpha$. The blue curve represents the real part of the Fourier transform, the orange curve depicts the imaginary part.}
\end{figure}

\begin{figure}[htbp] \label{experresults}
	\centering
	\includegraphics[width=\linewidth]{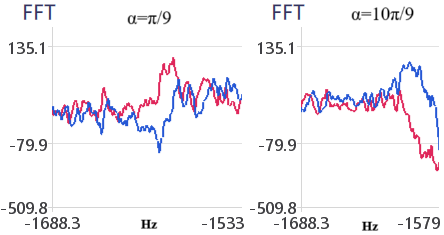}
	\caption{Experimental results of the NMR spectrum around frequencies close to the Larmor frequency of the qubit one, after applying Neural Network-generated pulses for the gates in Eq.\ref{eqalphateta} on pseudo-pure state for $\theta=\frac{\pi}{4}$ and represented $\alpha$. The blue curve represents the real part of the Fourier transform, the red curve depicts the imaginary part of the spectrum.}
\end{figure}

It is naturally expected that the experimental spectrum deviates from the theoretical one, as the theoretical model assumes a closed quantum system. In reality, however, the system is not closed. As a result, the pulse shapes derived from the theoretical model do not perform perfectly in practical scenarios, leading to discrepancies between the theoretical and experimental spectra under identical operations.
It is important to emphasize that quantum control and pulse shaping are employed to enhance gate fidelities by controlling unwanted interactions within closed quantum systems. However, these techniques do not address errors arising from environmental interactions and external sources of decoherence in open quantum systems. Several environmental factors contribute to the deviations observed between the predicted and measured spectra. These include non-homogeneous static magnetic fields, temporal fluctuations in magnetic field strength due to temperature variations, stray electromagnetic fields, and phase or amplitude noise in the RF signals.
For such systems, solutions like quantum error correction and error mitigation are required.
As discussed in the introduction, in closed quantum systems—particularly in NMR systems—the dominant sources of error include unwanted interactions of RF pulses with non-target qubits (especially in weak-field benchtop NMR systems) and unavoidable two-qubit interactions during single-qubit gate operations. In these cases, pulse shapes are engineered so that the target qubit undergoes the intended operation, while non-target qubits also evolve but return to their initial states by the end of the gate operation. In other words, the non-target qubits effectively experience an identity operation.

To demonstrate the consistency between the experimental results and the simulations, let us explicitly consider applying the gate defined in Eq.~\ref{eqalphateta} to the initial state \( \ket{000} \) which is equivalent to the pseudo-pure state:

\begin{equation}\label{gatedstate}
\begin{aligned}
e^{-i\frac{\pi}{4}(\cos(\alpha)\sigma_x^{(1)}+\sin(\alpha)\sigma_y^{(1)})}\ket{000} = \\
\frac{1}{\sqrt{2}}\left(\ket{000} + (-i\cos(\alpha) + \sin(\alpha))\ket{100}\right).
\end{aligned}
\end{equation}

Next, we consider the NMR signal spectrum corresponding to the above state and perform integration over its real and imaginary components within the frequency range associated with qubit~1:

\begin{equation}
\text{Im} = \int_{f_1}^{f_2} \text{im}(f) \, df, \quad 
\text{Re} = \int_{f_1}^{f_2} \text{re}(f) \, df,
\end{equation}
where \( f_1 \) to \( f_2 \) denotes the frequency range that exclusively contains the Larmor frequency of the target spin (qubit~1), excluding contributions from other spins (qubits~2 and~3). This selective integration allows us to isolate and quantify the signal intensity associated with qubit~1. The functions \( \text{im}(f) \) and \( \text{re}(f) \) represent the imaginary and real parts of the NMR spectrum, respectively.

We then define the phase of the output signal as

\begin{equation}\label{phasedef}
\text{Phase} = \tan^{-1}\left(\frac{\text{Im}}{\text{Re}}\right).
\end{equation}

By calculating the output NMR signal and its Fourier transform, as detailed in Appendix \ref{appendix}, we find that the quantities $Im$, $Re$ and $Phase$ for the state given in Eq.~\ref{gatedstate}, evaluated around the frequency corresponding to qubit 1, yield the following results:

\begin{equation} \label{derivationn}
\begin{aligned}
\text{Im} = -\cos(\alpha), \quad \text{Re} = \sin(\alpha)\\
\quad \Rightarrow \quad 
\text{Phase} = \tan^{-1}(-\cot(\alpha)),
\end{aligned}
\end{equation}
which demonstrates a linear relationship between the defined phase in Eq.~\ref{phasedef} and the parameter \( \alpha \) in Eq.~\ref{gatedstate}.

Finally, by obtaining the experimental NMR output spectrum corresponding to the state defined in Eq.~\ref{gatedstate} for various values of \( \alpha \) we compute the phase as described in Eq.~\ref{phasedef} for each value of \( \alpha \). The results are presented in Fig.14, alongside the theoretical predictions obtained from simulations. The close agreement between the experimental data and simulated results indicates that the model exhibits a good degree of accuracy at this scale, and that the neural network–generated pulses demonstrate reliable performance.

In the procedure of tomography in an NMR setting, we integrate the Fourier transform of the output. The relationship between the real and imaginary parts is crucial for recognizing the prepared state and applied quantum logic gates \cite{gg, a, b}. This test demonstrates the power of neural network-generated pulses in designing relevant quantum logic gates.

\begin{figure}[htbp]
	\centering
	\includegraphics[scale=0.4]{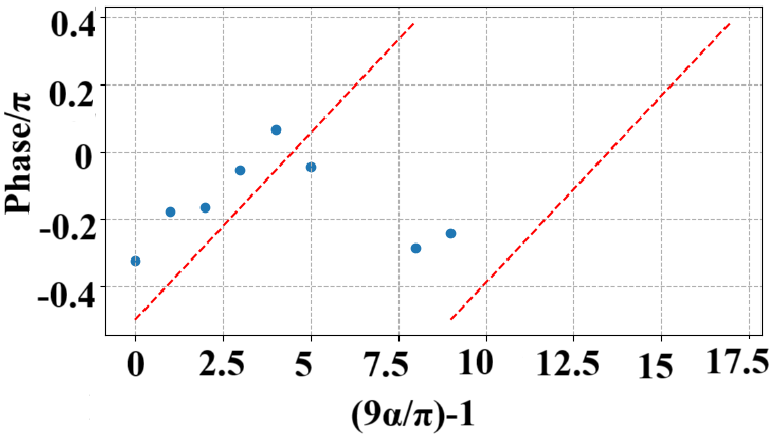}
	\caption{The phase defined in Eq.10 is plotted against $\alpha$ with $\theta=\frac{\pi}{4}$ for operation presented in Eq.6. The simulation data are represented by red lines and dashes, the experimental data are depicted with blue dots. This phase is measured for $\alpha=\frac{k\pi}{9}$ for $k \in \{1,2,3,4,5,6,9,10\}$.}
\end{figure}

\section{Conclusion}
In summary, we demonstrated the feasibility of extending optimal control theory for implementing arbitrary quantum logic gates through neural networks and illustrated the employment of the GRAPE algorithm to train the network. We demonstrated that using this approach, the cosine similarity between GRAPE-generated pulses used as training samples increases when the GRAPE algorithm is initiated from the same point, facilitating the learning process. After implementing the network, we tested it using numerical simulations, thereby verifying the success of the method. The feasibility of directly implementing arbitrary single-qubit gates without decomposing them into basic gates, helps achieve logical gates with higher fidelity and fewer steps. This approach significantly enhances circuit depth, which is particularly crucial in the NISQ era. When combined with the GRAPE-generated pulse for the CNOT gate, it creates an effective set of universal gates capable of implementing more complex algorithms in today's quantum computers. Moreover, the approximate real-time generation of control pulses for arbitrary operations allows the network to be integrated into the compiler within quantum computer architecture. Finally, we experimentally validated the effectiveness of the output control pulses generated by the developed neural network. For future research, further investigation can be carried out to understand the nature of the dips presented in Figs.8 and 9. Utilizing different architectures, data generation procedures, and physics-informed losses could potentially lead to higher fidelities and mitigate the network’s bias toward gates that were not sufficiently learned. It is also important to expand this concept to arbitrary two-qubit gates (i.e., Controlled-U gates) and explore the feasibility of designing a neural network for a larger number of qubits. This is crucial because, as the number of qubits increases, the system's dimensionality grows exponentially. This growth could hinder our ability to compute such large-dimensional matrices, rendering the calculation of overall dynamics infeasible. 

\appendix
\section{Derivation of Eq.\ref{derivationn}}
\label{appendix}
To derive Eq.\ref{derivationn}, we provide a more detailed discussion of the NMR spectrum. The output NMR signal is proportional to the magnetization in the plane perpendicular to the static magnetic field $B_0$ \cite{gg}:
\begin{equation}\label{Mx}
\begin{aligned}
\braket{M_x}=Tr(\rho(t)\sum^{n}_{k=1}\sigma^{(k)}_{x}),
\end{aligned}
\end{equation}
where $\rho(t)$ is the density matrix of the system at time $t$, during the free evolution of the system governed by the Hamiltonian $H_{free}$ in Eq.2. Using this, we obtain the following expression for $\braket{M_x}$:
\begin{equation}\label{MxSigma}
\begin{aligned}
Tr(e^{\frac{-it}{\hbar}H_{free}}\rho_0e^{\frac{it}{\hbar}H_{free}}\sum^{n}_{k=1}\sigma^{(k)}_{x}).
\end{aligned}
\end{equation}
Expanding $\rho_0$, the initial state of the system, in the eigenstates of $H_{free}$, we obtain:

\begin{equation}\label{magnet}
\begin{aligned} 
\mathrm{Tr}\left(e^{\frac{-it}{\hbar}H_{\text{free}}} \sum_{l,j} R_{l,j} \ket{l}\bra{j} 
e^{\frac{it}{\hbar}H_{\text{free}}} \sum_{k=1}^{n} \sigma^{(k)}_{x} \right) 
&= \nonumber \\
\sum_{l,j,k} R_{l,j} e^{-it(E_l - E_j)} \bra{j} \sigma^{(k)}_{x} \ket{l}, &
\end{aligned}
\end{equation}
where $E_l$ values are eigenvalues of $H_{free}$ and when we expanded $\rho_0$, we only considered the traceless part in Eq.\ref{gatedstate}, because the other part plays no role in the spectrum. Consequently, by substituting Eq.\ref{gatedstate} in the Eq.\ref{MxSigma}, we obtain the magnetization signal as:

\begin{equation}
\begin{aligned} 
\frac{1}{2} \left( -i e^{i\alpha} e^{\frac{-it}{\hbar}(E_{000} - E_{100})}
+ i e^{-i\alpha} e^{\frac{it}{\hbar}(E_{000} - E_{100})} \right)
= \nonumber \\
\mathrm{Re} \left( -i e^{i\alpha} e^{-i\omega t} \right)
= \sin(\alpha) \cos(\omega t) - \cos(\alpha) \sin(\omega t),
\end{aligned}
\end{equation}
where,
\begin{equation}
\omega=\frac{1}{\hbar}(E_{000}-E_{100}),
\end{equation}
By considering the resultant Fourier transform of the above signal, it is evident that the real part of the transform equals $\sin(\alpha)\delta(v-\omega)$, while the imaginary part equals -$\cos(\alpha)\delta(v-\omega)$. This approach allows us to demonstrate the validity of Eq.\ref{derivationn}.
Under nonideal conditions and in real experiments, some form of damping occurs in Eq.\ref{MxSigma} \cite{a}. This means that we can estimate the final signal as:

\begin{equation}\label{specdamped}
\begin{aligned}
e^{-t/T}(\sin(\alpha)\cos(\omega t)-\cos(\alpha)\sin(\omega t)).
\end{aligned}
\end{equation}

Again, by applying the Fourier transform to Eq.\ref{specdamped}, along with the detailed discussion in quantum state tomography on an NMR system \cite{xgg, gg, a, b}, we can demonstrate that the terms defined in Eq.\ref{MxSigma} justify Eq.\ref{derivationn}. By integrating the transformed signal, we obtain results that closely resemble the ideal situation.

\bibliography{NN}

\begin{thebibliography}{51}%
\makeatletter
\providecommand \@ifxundefined [1]{%
 \@ifx{#1\undefined}
}%
\providecommand \@ifnum [1]{%
 \ifnum #1\expandafter \@firstoftwo
 \else \expandafter \@secondoftwo
 \fi
}%
\providecommand \@ifx [1]{%
 \ifx #1\expandafter \@firstoftwo
 \else \expandafter \@secondoftwo
 \fi
}%
\providecommand \natexlab [1]{#1}%
\providecommand \enquote  [1]{``#1''}%
\providecommand \bibnamefont  [1]{#1}%
\providecommand \bibfnamefont [1]{#1}%
\providecommand \citenamefont [1]{#1}%
\providecommand \href@noop [0]{\@secondoftwo}%
\providecommand \href [0]{\begingroup \@sanitize@url \@href}%
\providecommand \@href[1]{\@@startlink{#1}\@@href}%
\providecommand \@@href[1]{\endgroup#1\@@endlink}%
\providecommand \@sanitize@url [0]{\catcode `\\12\catcode `\$12\catcode
  `\&12\catcode `\#12\catcode `\^12\catcode `\_12\catcode `\%12\relax}%
\providecommand \@@startlink[1]{}%
\providecommand \@@endlink[0]{}%
\providecommand \url  [0]{\begingroup\@sanitize@url \@url }%
\providecommand \@url [1]{\endgroup\@href {#1}{\urlprefix }}%
\providecommand \urlprefix  [0]{URL }%
\providecommand \Eprint [0]{\href }%
\providecommand \doibase [0]{https://doi.org/}%
\providecommand \selectlanguage [0]{\@gobble}%
\providecommand \bibinfo  [0]{\@secondoftwo}%
\providecommand \bibfield  [0]{\@secondoftwo}%
\providecommand \translation [1]{[#1]}%
\providecommand \BibitemOpen [0]{}%
\providecommand \bibitemStop [0]{}%
\providecommand \bibitemNoStop [0]{.\EOS\space}%
\providecommand \EOS [0]{\spacefactor3000\relax}%
\providecommand \BibitemShut  [1]{\csname bibitem#1\endcsname}%
\let\auto@bib@innerbib\@empty
\bibitem [{\citenamefont {Chang}\ \emph {et~al.}(2000)\citenamefont {Chang},
  \citenamefont {Vandersypen},\ and\ \citenamefont {Steffen}}]{a1}%
  \BibitemOpen
  \bibfield  {author} {\bibinfo {author} {\bibfnamefont {D.}~\bibnamefont
  {Chang}}, \bibinfo {author} {\bibfnamefont {L.}~\bibnamefont {Vandersypen}},\
  and\ \bibinfo {author} {\bibfnamefont {M.}~\bibnamefont {Steffen}},\
  }\bibfield  {title} {\bibinfo {title} {Nmr implementation of a building block
  for scalable quantum computation},\ }\href@noop {} {\bibfield  {journal}
  {\bibinfo  {journal} {Chemical Physics Letters}\ }\textbf {\bibinfo {volume}
  {338}},\ \bibinfo {pages} {337} (\bibinfo {year} {2000})}\BibitemShut
  {NoStop}%
\bibitem [{\citenamefont {Jones}(2024)}]{a}%
  \BibitemOpen
  \bibfield  {author} {\bibinfo {author} {\bibfnamefont {J.~A.}\ \bibnamefont
  {Jones}},\ }\bibfield  {title} {\bibinfo {title} {Controlling nmr spin
  systems for quantum computation},\ }\href@noop {} {\bibfield  {journal}
  {\bibinfo  {journal} {Progress in Nuclear Magnetic Resonance Spectroscopy}\
  }\textbf {\bibinfo {volume} {140–141}},\ \bibinfo {pages} {49} (\bibinfo
  {year} {2024})}\BibitemShut {NoStop}%
\bibitem [{\citenamefont {Jones}(2011)}]{b}%
  \BibitemOpen
  \bibfield  {author} {\bibinfo {author} {\bibfnamefont {J.~A.}\ \bibnamefont
  {Jones}},\ }\bibfield  {title} {\bibinfo {title} {Quantum computing with
  nmr},\ }\href@noop {} {\bibfield  {journal} {\bibinfo  {journal} {Progress in
  Nuclear Magnetic Resonance Spectroscopy}\ }\textbf {\bibinfo {volume}
  {59,2}},\ \bibinfo {pages} {91} (\bibinfo {year} {2011})}\BibitemShut
  {NoStop}%
\bibitem [{\citenamefont {Khaneja}\ \emph {et~al.}(2005)\citenamefont
  {Khaneja}, \citenamefont {Timo~Reiss}, \citenamefont {Schulte-Herbrüggen},\
  and\ \citenamefont {J.Glaser}}]{c}%
  \BibitemOpen
  \bibfield  {author} {\bibinfo {author} {\bibfnamefont {N.}~\bibnamefont
  {Khaneja}}, \bibinfo {author} {\bibfnamefont {C.~K.}\ \bibnamefont
  {Timo~Reiss}}, \bibinfo {author} {\bibfnamefont {T.}~\bibnamefont
  {Schulte-Herbrüggen}},\ and\ \bibinfo {author} {\bibfnamefont
  {S.}~\bibnamefont {J.Glaser}},\ }\bibfield  {title} {\bibinfo {title}
  {Optimal control of coupled spin dynamics: design of nmr pulse sequences by
  gradient ascent algorithms},\ }\href@noop {} {\bibfield  {journal} {\bibinfo
  {journal} {Journal of Magnetic Resonance}\ }\textbf {\bibinfo {volume}
  {172,2}},\ \bibinfo {pages} {296} (\bibinfo {year} {2005})}\BibitemShut
  {NoStop}%
\bibitem [{\citenamefont {Rasulov}\ \emph {et~al.}(2023)\citenamefont
  {Rasulov}, \citenamefont {Acharya}, \citenamefont {Carravetta}, \citenamefont
  {Mathies},\ and\ \citenamefont {Kuprov}}]{d}%
  \BibitemOpen
  \bibfield  {author} {\bibinfo {author} {\bibfnamefont {U.}~\bibnamefont
  {Rasulov}}, \bibinfo {author} {\bibfnamefont {A.}~\bibnamefont {Acharya}},
  \bibinfo {author} {\bibfnamefont {M.}~\bibnamefont {Carravetta}}, \bibinfo
  {author} {\bibfnamefont {G.}~\bibnamefont {Mathies}},\ and\ \bibinfo {author}
  {\bibfnamefont {I.}~\bibnamefont {Kuprov}},\ }\bibfield  {title} {\bibinfo
  {title} {Simulation and design of shaped pulses beyond the piecewise-constant
  approximation},\ }\href@noop {} {\bibfield  {journal} {\bibinfo  {journal}
  {Journal of Magnetic Resonance}\ }\textbf {\bibinfo {volume} {353}},\
  \bibinfo {pages} {107478} (\bibinfo {year} {2023})}\BibitemShut {NoStop}%
\bibitem [{\citenamefont {Peterson}\ \emph {et~al.}(2020)\citenamefont
  {Peterson}, \citenamefont {Sarthour},\ and\ \citenamefont {Laflamme}}]{q}%
  \BibitemOpen
  \bibfield  {author} {\bibinfo {author} {\bibfnamefont {J.~P.}\ \bibnamefont
  {Peterson}}, \bibinfo {author} {\bibfnamefont {R.~S.}\ \bibnamefont
  {Sarthour}},\ and\ \bibinfo {author} {\bibfnamefont {R.}~\bibnamefont
  {Laflamme}},\ }\bibfield  {title} {\bibinfo {title} {Enhancing quantum
  control by improving shaped-pulse generation},\ }\href@noop {} {\bibfield
  {journal} {\bibinfo  {journal} {PHYSICAL REVIEW D}\ } (\bibinfo {year}
  {2020})}\BibitemShut {NoStop}%
\bibitem [{\citenamefont {Kuzmanović}\ \emph {et~al.}(2024)\citenamefont
  {Kuzmanović}, \citenamefont {Björkman}, \citenamefont {McCord},
  \citenamefont {Dogra},\ and\ \citenamefont {Paraoanu}}]{y}%
  \BibitemOpen
  \bibfield  {author} {\bibinfo {author} {\bibfnamefont {M.}~\bibnamefont
  {Kuzmanović}}, \bibinfo {author} {\bibfnamefont {I.}~\bibnamefont
  {Björkman}}, \bibinfo {author} {\bibfnamefont {J.~J.}\ \bibnamefont
  {McCord}}, \bibinfo {author} {\bibfnamefont {S.}~\bibnamefont {Dogra}},\ and\
  \bibinfo {author} {\bibfnamefont {G.~S.}\ \bibnamefont {Paraoanu}},\
  }\bibfield  {title} {\bibinfo {title} {High-fidelity robust qubit control by
  phase-modulated pulses},\ }\href@noop {} {\bibfield  {journal} {\bibinfo
  {journal} {PHYSICAL REVIEW RESEARCH}\ } (\bibinfo {year} {2024})}\BibitemShut
  {NoStop}%
\bibitem [{\citenamefont {Ryan}\ \emph {et~al.}(2008)\citenamefont {Ryan},
  \citenamefont {Negrevergne}, \citenamefont {Laforest}, \citenamefont
  {Knill},\ and\ \citenamefont {Laflamme}}]{cc}%
  \BibitemOpen
  \bibfield  {author} {\bibinfo {author} {\bibfnamefont {C.~A.}\ \bibnamefont
  {Ryan}}, \bibinfo {author} {\bibfnamefont {C.}~\bibnamefont {Negrevergne}},
  \bibinfo {author} {\bibfnamefont {M.}~\bibnamefont {Laforest}}, \bibinfo
  {author} {\bibfnamefont {E.}~\bibnamefont {Knill}},\ and\ \bibinfo {author}
  {\bibfnamefont {R.}~\bibnamefont {Laflamme}},\ }\bibfield  {title} {\bibinfo
  {title} {Liquid state nmr as a test-bed for developing quantum control
  methods},\ }\href@noop {} {\bibfield  {journal} {\bibinfo  {journal} {NIST}\
  } (\bibinfo {year} {2008})}\BibitemShut {NoStop}%
\bibitem [{\citenamefont {Chen}\ \emph {et~al.}(2023)\citenamefont {Chen},
  \citenamefont {Hao}, \citenamefont {Wu}, \citenamefont {Wang}, \citenamefont
  {Liu}, \citenamefont {Hou}, \citenamefont {Cui}, \citenamefont {Yung}, ,\
  and\ \citenamefont {Peng}}]{e}%
  \BibitemOpen
  \bibfield  {author} {\bibinfo {author} {\bibfnamefont {Y.}~\bibnamefont
  {Chen}}, \bibinfo {author} {\bibfnamefont {Y.}~\bibnamefont {Hao}}, \bibinfo
  {author} {\bibfnamefont {Z.}~\bibnamefont {Wu}}, \bibinfo {author}
  {\bibfnamefont {B.-Y.}\ \bibnamefont {Wang}}, \bibinfo {author}
  {\bibfnamefont {R.}~\bibnamefont {Liu}}, \bibinfo {author} {\bibfnamefont
  {Y.}~\bibnamefont {Hou}}, \bibinfo {author} {\bibfnamefont {J.}~\bibnamefont
  {Cui}}, \bibinfo {author} {\bibfnamefont {M.-H.}\ \bibnamefont {Yung}}, ,\
  and\ \bibinfo {author} {\bibfnamefont {X.}~\bibnamefont {Peng}},\ }\bibfield
  {title} {\bibinfo {title} {Accelerating quantum optimal control through
  iterative gradient-ascent pulse engineering},\ }\href@noop {} {\bibfield
  {journal} {\bibinfo  {journal} {PHYSICAL REVIEW A}\ }\textbf {\bibinfo
  {volume} {108}} (\bibinfo {year} {2023})}\BibitemShut {NoStop}%
\bibitem [{\citenamefont {Ram}\ \emph {et~al.}(2022)\citenamefont {Ram},
  \citenamefont {Krithika}, \citenamefont {Batra}, ,\ and\ \citenamefont
  {Mahesh}}]{f}%
  \BibitemOpen
  \bibfield  {author} {\bibinfo {author} {\bibfnamefont {M.~H.}\ \bibnamefont
  {Ram}}, \bibinfo {author} {\bibfnamefont {V.~R.}\ \bibnamefont {Krithika}},
  \bibinfo {author} {\bibfnamefont {P.}~\bibnamefont {Batra}}, ,\ and\ \bibinfo
  {author} {\bibfnamefont {T.~S.}\ \bibnamefont {Mahesh}},\ }\bibfield  {title}
  {\bibinfo {title} {Robust quantum control using hybrid pulse engineering},\
  }\href@noop {} {\bibfield  {journal} {\bibinfo  {journal} {PHYSICAL REVIEW
  A}\ }\textbf {\bibinfo {volume} {105}},\ \bibinfo {pages} {042437} (\bibinfo
  {year} {2022})}\BibitemShut {NoStop}%
\bibitem [{\citenamefont {Lia}\ \emph {et~al.}(2017)\citenamefont {Lia},
  \citenamefont {Yang}, \citenamefont {Peng},\ and\ \citenamefont {Sun}}]{aa}%
  \BibitemOpen
  \bibfield  {author} {\bibinfo {author} {\bibfnamefont {J.}~\bibnamefont
  {Lia}}, \bibinfo {author} {\bibfnamefont {X.}~\bibnamefont {Yang}}, \bibinfo
  {author} {\bibfnamefont {X.}~\bibnamefont {Peng}},\ and\ \bibinfo {author}
  {\bibfnamefont {C.-P.}\ \bibnamefont {Sun}},\ }\bibfield  {title} {\bibinfo
  {title} {Hybrid quantum-classical approach to quantum optimal control},\
  }\href@noop {} {\bibfield  {journal} {\bibinfo  {journal} {PHYSICAL REVIEW
  LETTERS}\ } (\bibinfo {year} {2017})}\BibitemShut {NoStop}%
\bibitem [{\citenamefont {dong Yang}\ \emph {et~al.}(2020)\citenamefont {dong
  Yang}, \citenamefont {Arenz}, \citenamefont {Pelczer}, \citenamefont {Chen},
  \citenamefont {Wu}, \citenamefont {Peng},\ and\ \citenamefont {Rabitz}}]{bb}%
  \BibitemOpen
  \bibfield  {author} {\bibinfo {author} {\bibfnamefont {X.}~\bibnamefont {dong
  Yang}}, \bibinfo {author} {\bibfnamefont {C.}~\bibnamefont {Arenz}}, \bibinfo
  {author} {\bibfnamefont {I.}~\bibnamefont {Pelczer}}, \bibinfo {author}
  {\bibfnamefont {Q.-M.}\ \bibnamefont {Chen}}, \bibinfo {author}
  {\bibfnamefont {R.-B.}\ \bibnamefont {Wu}}, \bibinfo {author} {\bibfnamefont
  {X.}~\bibnamefont {Peng}},\ and\ \bibinfo {author} {\bibfnamefont
  {H.}~\bibnamefont {Rabitz}},\ }\bibfield  {title} {\bibinfo {title}
  {Assessing three closed-loop learning algorithms by searching for
  high-quality quantum control pulses},\ }\href@noop {} {\bibfield  {journal}
  {\bibinfo  {journal} {PHYSICAL REVIEW A}\ } (\bibinfo {year}
  {2020})}\BibitemShut {NoStop}%
\bibitem [{\citenamefont {Kairys}\ and\ \citenamefont {Humble}(2021)}]{dd}%
  \BibitemOpen
  \bibfield  {author} {\bibinfo {author} {\bibfnamefont {P.}~\bibnamefont
  {Kairys}}\ and\ \bibinfo {author} {\bibfnamefont {T.~S.}\ \bibnamefont
  {Humble}},\ }\bibfield  {title} {\bibinfo {title} {Efficient quantum gate
  discovery with optimal control},\ }\href@noop {} {\bibfield  {journal}
  {\bibinfo  {journal} {IEEE International Conference on Quantum Computing and
  Engineering (QCE)}\ } (\bibinfo {year} {2021})}\BibitemShut {NoStop}%
\bibitem [{\citenamefont {Riaz}\ \emph {et~al.}(2019)\citenamefont {Riaz},
  \citenamefont {Shuang},\ and\ \citenamefont {Qamar}}]{g}%
  \BibitemOpen
  \bibfield  {author} {\bibinfo {author} {\bibfnamefont {B.}~\bibnamefont
  {Riaz}}, \bibinfo {author} {\bibfnamefont {C.}~\bibnamefont {Shuang}},\ and\
  \bibinfo {author} {\bibfnamefont {S.}~\bibnamefont {Qamar}},\ }\bibfield
  {title} {\bibinfo {title} {Optimal control methods for quantum gate
  preparation: a comparative study},\ }\href@noop {} {\bibfield  {journal}
  {\bibinfo  {journal} {Quantum Information Processing}\ }\textbf {\bibinfo
  {volume} {18}} (\bibinfo {year} {2019})}\BibitemShut {NoStop}%
\bibitem [{\citenamefont {Goodwin}\ and\ \citenamefont {Kuprov}(2016)}]{h}%
  \BibitemOpen
  \bibfield  {author} {\bibinfo {author} {\bibfnamefont {D.~L.}\ \bibnamefont
  {Goodwin}}\ and\ \bibinfo {author} {\bibfnamefont {I.}~\bibnamefont
  {Kuprov}},\ }\bibfield  {title} {\bibinfo {title} {Modified newton-raphson
  grape methods for optimal control of spin systems},\ }\href@noop {}
  {\bibfield  {journal} {\bibinfo  {journal} {The journal of Chemical Physics}\
  }\textbf {\bibinfo {volume} {144}} (\bibinfo {year} {2016})}\BibitemShut
  {NoStop}%
\bibitem [{\citenamefont {Sauvage}\ and\ \citenamefont {Mintert}(2022)}]{i}%
  \BibitemOpen
  \bibfield  {author} {\bibinfo {author} {\bibfnamefont {F.}~\bibnamefont
  {Sauvage}}\ and\ \bibinfo {author} {\bibfnamefont {F.}~\bibnamefont
  {Mintert}},\ }\bibfield  {title} {\bibinfo {title} {Optimal control of
  families of quantum gates},\ }\href@noop {} {\bibfield  {journal} {\bibinfo
  {journal} {PHYSICAL REVIEW LETTERS}\ }\textbf {\bibinfo {volume} {129}}
  (\bibinfo {year} {2022})}\BibitemShut {NoStop}%
\bibitem [{\citenamefont {Ma}\ \emph {et~al.}(2022)\citenamefont {Ma},
  \citenamefont {Dong}, \citenamefont {Ding},\ and\ \citenamefont {Chen}}]{k}%
  \BibitemOpen
  \bibfield  {author} {\bibinfo {author} {\bibfnamefont {H.}~\bibnamefont
  {Ma}}, \bibinfo {author} {\bibfnamefont {D.}~\bibnamefont {Dong}}, \bibinfo
  {author} {\bibfnamefont {S.~X.}\ \bibnamefont {Ding}},\ and\ \bibinfo
  {author} {\bibfnamefont {C.}~\bibnamefont {Chen}},\ }\bibfield  {title}
  {\bibinfo {title} {Curriculum-based deep reinforcement learning for quantum
  control},\ }\href@noop {} {\bibfield  {journal} {\bibinfo  {journal} {IEEE
  Transactions on Neural Networks and Learning Systems}\ }\textbf {\bibinfo
  {volume} {34}} (\bibinfo {year} {2022})}\BibitemShut {NoStop}%
\bibitem [{\citenamefont {Zhang}\ \emph {et~al.}(2019)\citenamefont {Zhang},
  \citenamefont {Wei}, \citenamefont {Asad}, \citenamefont {Yang},\ and\
  \citenamefont {Wang}}]{l}%
  \BibitemOpen
  \bibfield  {author} {\bibinfo {author} {\bibfnamefont {X.-M.}\ \bibnamefont
  {Zhang}}, \bibinfo {author} {\bibfnamefont {Z.}~\bibnamefont {Wei}}, \bibinfo
  {author} {\bibfnamefont {R.}~\bibnamefont {Asad}}, \bibinfo {author}
  {\bibfnamefont {X.-C.}\ \bibnamefont {Yang}},\ and\ \bibinfo {author}
  {\bibfnamefont {X.}~\bibnamefont {Wang}},\ }\bibfield  {title} {\bibinfo
  {title} {When does reinforcement learning stand out in quantum control? a
  comparative study on state preparation},\ }\href@noop {} {\bibfield
  {journal} {\bibinfo  {journal} {npj Quantum Information volume}\ ,\ \bibinfo
  {pages} {5}} (\bibinfo {year} {2019})}\BibitemShut {NoStop}%
\bibitem [{\citenamefont {Güngördü}\ and\ \citenamefont
  {Kestner}(2022)}]{m}%
  \BibitemOpen
  \bibfield  {author} {\bibinfo {author} {\bibfnamefont {U.}~\bibnamefont
  {Güngördü}}\ and\ \bibinfo {author} {\bibfnamefont {J.~P.}\ \bibnamefont
  {Kestner}},\ }\bibfield  {title} {\bibinfo {title} {Robust quantum gates
  using smooth pulses and physics-informed neural networks},\ }\href@noop {}
  {\bibfield  {journal} {\bibinfo  {journal} {PHYSICAL REVIEW RESEARCH}\
  }\textbf {\bibinfo {volume} {4}} (\bibinfo {year} {2022})}\BibitemShut
  {NoStop}%
\bibitem [{\citenamefont {Norambuena}\ \emph {et~al.}(2023)\citenamefont
  {Norambuena}, \citenamefont {Mattheakis}, \citenamefont {González},\ and\
  \citenamefont {Coto}}]{n}%
  \BibitemOpen
  \bibfield  {author} {\bibinfo {author} {\bibfnamefont {A.}~\bibnamefont
  {Norambuena}}, \bibinfo {author} {\bibfnamefont {M.}~\bibnamefont
  {Mattheakis}}, \bibinfo {author} {\bibfnamefont {F.~J.}\ \bibnamefont
  {González}},\ and\ \bibinfo {author} {\bibfnamefont {R.}~\bibnamefont
  {Coto}},\ }\bibfield  {title} {\bibinfo {title} {Physics-informed neural
  networks for quantum control},\ }\href@noop {} {\bibfield  {journal}
  {\bibinfo  {journal} {arXiv:2206.06287}\ } (\bibinfo {year}
  {2023})}\BibitemShut {NoStop}%
\bibitem [{\citenamefont {Youssry}\ \emph {et~al.}(2024)\citenamefont
  {Youssry}, \citenamefont {Yang}, \citenamefont {Chapman}, \citenamefont
  {Haylock}, \citenamefont {Lenzini}, \citenamefont {Lobino},\ and\
  \citenamefont {Peruzzo}}]{o}%
  \BibitemOpen
  \bibfield  {author} {\bibinfo {author} {\bibfnamefont {A.}~\bibnamefont
  {Youssry}}, \bibinfo {author} {\bibfnamefont {Y.}~\bibnamefont {Yang}},
  \bibinfo {author} {\bibfnamefont {R.~J.}\ \bibnamefont {Chapman}}, \bibinfo
  {author} {\bibfnamefont {B.}~\bibnamefont {Haylock}}, \bibinfo {author}
  {\bibfnamefont {F.}~\bibnamefont {Lenzini}}, \bibinfo {author} {\bibfnamefont
  {M.}~\bibnamefont {Lobino}},\ and\ \bibinfo {author} {\bibfnamefont
  {A.}~\bibnamefont {Peruzzo}},\ }\bibfield  {title} {\bibinfo {title}
  {Experimental graybox quantum system identification and control},\
  }\href@noop {} {\bibfield  {journal} {\bibinfo  {journal} {npj Quantum
  Information volume}\ } (\bibinfo {year} {2024})}\BibitemShut {NoStop}%
\bibitem [{\citenamefont {Sivak}\ \emph {et~al.}(2022)\citenamefont {Sivak},
  \citenamefont {Eickbusch}, \citenamefont {H.~Liu}, \citenamefont
  {Tsioutsios},\ and\ \citenamefont {Devoret}}]{p}%
  \BibitemOpen
  \bibfield  {author} {\bibinfo {author} {\bibfnamefont {V.}~\bibnamefont
  {Sivak}}, \bibinfo {author} {\bibfnamefont {A.}~\bibnamefont {Eickbusch}},
  \bibinfo {author} {\bibfnamefont {B.~R.}\ \bibnamefont {H.~Liu}}, \bibinfo
  {author} {\bibfnamefont {I.}~\bibnamefont {Tsioutsios}},\ and\ \bibinfo
  {author} {\bibfnamefont {M.}~\bibnamefont {Devoret}},\ }\bibfield  {title}
  {\bibinfo {title} {Model-free quantum control with reinforcement learning},\
  }\href@noop {} {\bibfield  {journal} {\bibinfo  {journal} {PHYSICAL REVIEW
  X}\ } (\bibinfo {year} {2022})}\BibitemShut {NoStop}%
\bibitem [{\citenamefont {Wu}\ \emph {et~al.}(2016)\citenamefont {Wu},
  \citenamefont {Qi}, \citenamefont {Chen},\ and\ \citenamefont {Dong}}]{r}%
  \BibitemOpen
  \bibfield  {author} {\bibinfo {author} {\bibfnamefont {C.}~\bibnamefont
  {Wu}}, \bibinfo {author} {\bibfnamefont {B.}~\bibnamefont {Qi}}, \bibinfo
  {author} {\bibfnamefont {C.}~\bibnamefont {Chen}},\ and\ \bibinfo {author}
  {\bibfnamefont {D.}~\bibnamefont {Dong}},\ }\bibfield  {title} {\bibinfo
  {title} {Robust learning control design for quantum unitary
  transformations},\ }\href@noop {} {\bibfield  {journal} {\bibinfo  {journal}
  {IEEE Transactions on Cybernetics}\ }\textbf {\bibinfo {volume} {17}}
  (\bibinfo {year} {2016})}\BibitemShut {NoStop}%
\bibitem [{\citenamefont {Yang}\ \emph {et~al.}(2018)\citenamefont {Yang},
  \citenamefont {Yung},\ and\ \citenamefont {Wang}}]{t}%
  \BibitemOpen
  \bibfield  {author} {\bibinfo {author} {\bibfnamefont {X.-C.}\ \bibnamefont
  {Yang}}, \bibinfo {author} {\bibfnamefont {M.-H.}\ \bibnamefont {Yung}},\
  and\ \bibinfo {author} {\bibfnamefont {X.}~\bibnamefont {Wang}},\ }\bibfield
  {title} {\bibinfo {title} {Neural-network-designed pulse sequences for robust
  control of singlet-triplet qubits},\ }\href@noop {} {\bibfield  {journal}
  {\bibinfo  {journal} {PHYSICAL REVIEW A}\ }\textbf {\bibinfo {volume}
  {042324}} (\bibinfo {year} {2018})}\BibitemShut {NoStop}%
\bibitem [{\citenamefont {Wang}\ \emph {et~al.}(2020)\citenamefont {Wang},
  \citenamefont {Kumar}, \citenamefont {Shelton},\ and\ \citenamefont
  {Wong}}]{u}%
  \BibitemOpen
  \bibfield  {author} {\bibinfo {author} {\bibfnamefont {X.}~\bibnamefont
  {Wang}}, \bibinfo {author} {\bibfnamefont {A.}~\bibnamefont {Kumar}},
  \bibinfo {author} {\bibfnamefont {C.~R.}\ \bibnamefont {Shelton}},\ and\
  \bibinfo {author} {\bibfnamefont {B.~M.}\ \bibnamefont {Wong}},\ }\bibfield
  {title} {\bibinfo {title} {Harnessing deep neural networks to solve inverse
  problems in quantum dynamics: machine-learned predictions of time-dependent
  optimal control fields},\ }\href@noop {} {\bibfield  {journal} {\bibinfo
  {journal} {Physical Chemistry Chemical Physics}\ } (\bibinfo {year}
  {2020})}\BibitemShut {NoStop}%
\bibitem [{\citenamefont {He}\ \emph {et~al.}(2021)\citenamefont {He},
  \citenamefont {Nie}, \citenamefont {Wu}, \citenamefont {Zhang},\ and\
  \citenamefont {Wang}}]{v}%
  \BibitemOpen
  \bibfield  {author} {\bibinfo {author} {\bibfnamefont {R.-H.}\ \bibnamefont
  {He}}, \bibinfo {author} {\bibfnamefont {R.~W. S.-S.}\ \bibnamefont {Nie}},
  \bibinfo {author} {\bibfnamefont {J.}~\bibnamefont {Wu}}, \bibinfo {author}
  {\bibfnamefont {J.-H.}\ \bibnamefont {Zhang}},\ and\ \bibinfo {author}
  {\bibfnamefont {Z.-M.}\ \bibnamefont {Wang}},\ }\bibfield  {title} {\bibinfo
  {title} {Deep reinforcement learning for universal quantum state preparation
  via dynamic pulse control},\ }\href@noop {} {\bibfield  {journal} {\bibinfo
  {journal} {EPJ Quantum Technology}\ } (\bibinfo {year} {2021})}\BibitemShut
  {NoStop}%
\bibitem [{\citenamefont {Niu}\ \emph {et~al.}(2019)\citenamefont {Niu},
  \citenamefont {Boixo}, \citenamefont {Smelyanskiy},\ and\ \citenamefont
  {Neven}}]{w}%
  \BibitemOpen
  \bibfield  {author} {\bibinfo {author} {\bibfnamefont {M.~Y.}\ \bibnamefont
  {Niu}}, \bibinfo {author} {\bibfnamefont {S.}~\bibnamefont {Boixo}}, \bibinfo
  {author} {\bibfnamefont {V.~N.}\ \bibnamefont {Smelyanskiy}},\ and\ \bibinfo
  {author} {\bibfnamefont {H.}~\bibnamefont {Neven}},\ }\bibfield  {title}
  {\bibinfo {title} {Universal quantum control through deep reinforcement
  learning},\ }\href@noop {} {\bibfield  {journal} {\bibinfo  {journal} {npj
  Quantum Information}\ } (\bibinfo {year} {2019})}\BibitemShut {NoStop}%
\bibitem [{\citenamefont {Wu}\ \emph {et~al.}(2019)\citenamefont {Wu},
  \citenamefont {Ding}, \citenamefont {Dong},\ and\ \citenamefont {Wang}}]{x}%
  \BibitemOpen
  \bibfield  {author} {\bibinfo {author} {\bibfnamefont {R.-B.}\ \bibnamefont
  {Wu}}, \bibinfo {author} {\bibfnamefont {H.}~\bibnamefont {Ding}}, \bibinfo
  {author} {\bibfnamefont {D.}~\bibnamefont {Dong}},\ and\ \bibinfo {author}
  {\bibfnamefont {X.}~\bibnamefont {Wang}},\ }\bibfield  {title} {\bibinfo
  {title} {Learning robust and high-precision quantum controls},\ }\href@noop
  {} {\bibfield  {journal} {\bibinfo  {journal} {PHYSICAL REVIEW A}\ }
  (\bibinfo {year} {2019})}\BibitemShut {NoStop}%
\bibitem [{\citenamefont {Fösel}\ \emph {et~al.}(2018)\citenamefont {Fösel},
  \citenamefont {Tighineanu}, \citenamefont {Weiss},\ and\ \citenamefont
  {Marquardt}}]{z}%
  \BibitemOpen
  \bibfield  {author} {\bibinfo {author} {\bibfnamefont {T.}~\bibnamefont
  {Fösel}}, \bibinfo {author} {\bibfnamefont {P.}~\bibnamefont {Tighineanu}},
  \bibinfo {author} {\bibfnamefont {T.}~\bibnamefont {Weiss}},\ and\ \bibinfo
  {author} {\bibfnamefont {F.}~\bibnamefont {Marquardt}},\ }\bibfield  {title}
  {\bibinfo {title} {Reinforcement learning with neural networks for quantum
  feedback},\ }\href@noop {} {\bibfield  {journal} {\bibinfo  {journal}
  {PHYSICAL REVIEW X}\ } (\bibinfo {year} {2018})}\BibitemShut {NoStop}%
\bibitem [{\citenamefont {Jang}\ \emph {et~al.}(2024)\citenamefont {Jang},
  \citenamefont {He},\ and\ \citenamefont {Liu}}]{ii}%
  \BibitemOpen
  \bibfield  {author} {\bibinfo {author} {\bibfnamefont {A.}~\bibnamefont
  {Jang}}, \bibinfo {author} {\bibfnamefont {X.}~\bibnamefont {He}},\ and\
  \bibinfo {author} {\bibfnamefont {F.}~\bibnamefont {Liu}},\ }\bibfield
  {title} {\bibinfo {title} {Physics-guided self-supervised learning:
  Demonstration for generalized rf pulse design},\ }\href@noop {} {\bibfield
  {journal} {\bibinfo  {journal} {Magnetic Resonance in Medicine}\ }\textbf
  {\bibinfo {volume} {93}},\ \bibinfo {pages} {657} (\bibinfo {year}
  {2024})}\BibitemShut {NoStop}%
\bibitem [{\citenamefont {Zhang}\ \emph {et~al.}(2021)\citenamefont {Zhang},
  \citenamefont {Jiang}, \citenamefont {Jiang}, \citenamefont {Wang},
  \citenamefont {Wright}, \citenamefont {Liu},\ and\ \citenamefont
  {Wang}}]{jj}%
  \BibitemOpen
  \bibfield  {author} {\bibinfo {author} {\bibfnamefont {Y.}~\bibnamefont
  {Zhang}}, \bibinfo {author} {\bibfnamefont {K.}~\bibnamefont {Jiang}},
  \bibinfo {author} {\bibfnamefont {W.}~\bibnamefont {Jiang}}, \bibinfo
  {author} {\bibfnamefont {N.}~\bibnamefont {Wang}}, \bibinfo {author}
  {\bibfnamefont {A.~J.}\ \bibnamefont {Wright}}, \bibinfo {author}
  {\bibfnamefont {A.}~\bibnamefont {Liu}},\ and\ \bibinfo {author}
  {\bibfnamefont {J.}~\bibnamefont {Wang}},\ }\bibfield  {title} {\bibinfo
  {title} {Multi-task convolutional neural network-based design of radio
  frequency pulse and the accompanying gradients for magnetic resonance
  imaging},\ }\href@noop {} {\bibfield  {journal} {\bibinfo  {journal} {NMR in
  Biomedicine}\ }\textbf {\bibinfo {volume} {34}},\ \bibinfo {pages} {e4443}
  (\bibinfo {year} {2021})}\BibitemShut {NoStop}%
\bibitem [{\citenamefont {Gezelter}\ and\ \citenamefont {Freeman}(1990)}]{kk}%
  \BibitemOpen
  \bibfield  {author} {\bibinfo {author} {\bibfnamefont {J.}~\bibnamefont
  {Gezelter}}\ and\ \bibinfo {author} {\bibfnamefont {R.}~\bibnamefont
  {Freeman}},\ }\bibfield  {title} {\bibinfo {title} {Use of neural networks to
  design shaped radiofrequency pulses},\ }\href
  {https://doi.org/https://doi.org/10.1016/0022-2364(90)90149-4} {\bibfield
  {journal} {\bibinfo  {journal} {Journal of Magnetic Resonance (1969)}\
  }\textbf {\bibinfo {volume} {90}},\ \bibinfo {pages} {397} (\bibinfo {year}
  {1990})}\BibitemShut {NoStop}%
\bibitem [{\citenamefont {Vinding}\ and\ \citenamefont {Lund}(2023)}]{ll}%
  \BibitemOpen
  \bibfield  {author} {\bibinfo {author} {\bibfnamefont {M.~S.}\ \bibnamefont
  {Vinding}}\ and\ \bibinfo {author} {\bibfnamefont {T.~E.}\ \bibnamefont
  {Lund}},\ }\bibfield  {title} {\bibinfo {title} {Clipped deepcontrol: Deep
  neural network two-dimensional pulse design with an amplitude constraint
  layer},\ }\href@noop {} {\bibfield  {journal} {\bibinfo  {journal}
  {Artificial Intelligence in Medicine}\ }\textbf {\bibinfo {volume} {135}},\
  \bibinfo {pages} {102460} (\bibinfo {year} {2023})}\BibitemShut {NoStop}%
\bibitem [{\citenamefont {Tokarz}(2024)}]{mm}%
  \BibitemOpen
  \bibfield  {author} {\bibinfo {author} {\bibfnamefont {P.}~\bibnamefont
  {Tokarz}},\ }\bibfield  {title} {\bibinfo {title} {Artificial
  intelligence-powered pulse sequences in nuclear magnetic resonance and
  magnetic resonance imaging: Historical trends, current innovations and
  perspectives},\ }\href@noop {} {\bibfield  {journal} {\bibinfo  {journal}
  {Scientiae Radices}\ } (\bibinfo {year} {2024})}\BibitemShut {NoStop}%
\bibitem [{\citenamefont {Vinding}\ \emph {et~al.}(2019)\citenamefont
  {Vinding}, \citenamefont {Skyum}, \citenamefont {Sangill},\ and\
  \citenamefont {Lund}}]{nn}%
  \BibitemOpen
  \bibfield  {author} {\bibinfo {author} {\bibfnamefont {M.~S.}\ \bibnamefont
  {Vinding}}, \bibinfo {author} {\bibfnamefont {B.}~\bibnamefont {Skyum}},
  \bibinfo {author} {\bibfnamefont {R.}~\bibnamefont {Sangill}},\ and\ \bibinfo
  {author} {\bibfnamefont {T.~E.}\ \bibnamefont {Lund}},\ }\bibfield  {title}
  {\bibinfo {title} {Ultrafast (milliseconds), multidimensional rf pulse design
  with deep learning},\ }\href@noop {} {\bibfield  {journal} {\bibinfo
  {journal} {Magnetic Resonance in Medicine}\ }\textbf {\bibinfo {volume}
  {82}},\ \bibinfo {pages} {586} (\bibinfo {year} {2019})}\BibitemShut
  {NoStop}%
\bibitem [{\citenamefont {Vinding}\ \emph {et~al.}(2021)\citenamefont
  {Vinding}, \citenamefont {Aigner}, \citenamefont {Schmitter},\ and\
  \citenamefont {Lund}}]{oo}%
  \BibitemOpen
  \bibfield  {author} {\bibinfo {author} {\bibfnamefont {M.~S.}\ \bibnamefont
  {Vinding}}, \bibinfo {author} {\bibfnamefont {C.~S.}\ \bibnamefont {Aigner}},
  \bibinfo {author} {\bibfnamefont {S.}~\bibnamefont {Schmitter}},\ and\
  \bibinfo {author} {\bibfnamefont {T.~E.}\ \bibnamefont {Lund}},\ }\bibfield
  {title} {\bibinfo {title} {Deepcontrol: 2drf pulses facilitating
  inhomogeneity and b0 off-resonance compensation in vivo at 7 t},\ }\href@noop
  {} {\bibfield  {journal} {\bibinfo  {journal} {Magnetic Resonance in
  Medicine}\ }\textbf {\bibinfo {volume} {85}},\ \bibinfo {pages} {3308}
  (\bibinfo {year} {2021})}\BibitemShut {NoStop}%
\bibitem [{\citenamefont {Tomi-Tricot}\ \emph {et~al.}(2019)\citenamefont
  {Tomi-Tricot}, \citenamefont {Gras}, \citenamefont {Thirion}, \citenamefont
  {Mauconduit}, \citenamefont {Boulant}, \citenamefont {Cherkaoui},
  \citenamefont {Zerbib}, \citenamefont {Vignaud}, \citenamefont {Luciani},\
  and\ \citenamefont {Amadon}}]{pp}%
  \BibitemOpen
  \bibfield  {author} {\bibinfo {author} {\bibfnamefont {R.}~\bibnamefont
  {Tomi-Tricot}}, \bibinfo {author} {\bibfnamefont {V.}~\bibnamefont {Gras}},
  \bibinfo {author} {\bibfnamefont {B.}~\bibnamefont {Thirion}}, \bibinfo
  {author} {\bibfnamefont {F.}~\bibnamefont {Mauconduit}}, \bibinfo {author}
  {\bibfnamefont {N.}~\bibnamefont {Boulant}}, \bibinfo {author} {\bibfnamefont
  {H.}~\bibnamefont {Cherkaoui}}, \bibinfo {author} {\bibfnamefont
  {P.}~\bibnamefont {Zerbib}}, \bibinfo {author} {\bibfnamefont
  {A.}~\bibnamefont {Vignaud}}, \bibinfo {author} {\bibfnamefont
  {A.}~\bibnamefont {Luciani}},\ and\ \bibinfo {author} {\bibfnamefont
  {A.}~\bibnamefont {Amadon}},\ }\bibfield  {title} {\bibinfo {title}
  {Smartpulse, a machine learning approach for calibration-free dynamic rf
  shimming: Preliminary study in a clinical environment},\ }\href@noop {}
  {\bibfield  {journal} {\bibinfo  {journal} {Magnetic Resonance in Medicine}\
  }\textbf {\bibinfo {volume} {82}},\ \bibinfo {pages} {2016} (\bibinfo {year}
  {2019})}\BibitemShut {NoStop}%
\bibitem [{\citenamefont {Becker}\ \emph {et~al.}(2023)\citenamefont {Becker},
  \citenamefont {Cheng}, \citenamefont {Voigt}, \citenamefont {Chenakkara},
  \citenamefont {He}, \citenamefont {Lehmkuhl}, \citenamefont {Jouda},\ and\
  \citenamefont {Korvink}}]{qq}%
  \BibitemOpen
  \bibfield  {author} {\bibinfo {author} {\bibfnamefont {M.}~\bibnamefont
  {Becker}}, \bibinfo {author} {\bibfnamefont {Y.-T.}\ \bibnamefont {Cheng}},
  \bibinfo {author} {\bibfnamefont {A.}~\bibnamefont {Voigt}}, \bibinfo
  {author} {\bibfnamefont {A.}~\bibnamefont {Chenakkara}}, \bibinfo {author}
  {\bibfnamefont {M.}~\bibnamefont {He}}, \bibinfo {author} {\bibfnamefont
  {S.}~\bibnamefont {Lehmkuhl}}, \bibinfo {author} {\bibfnamefont
  {M.}~\bibnamefont {Jouda}},\ and\ \bibinfo {author} {\bibfnamefont {J.~G.}\
  \bibnamefont {Korvink}},\ }\bibfield  {title} {\bibinfo {title} {Artificial
  intelligence-driven shimming for parallel high field nuclear magnetic
  resonance},\ }\href@noop {} {\bibfield  {journal} {\bibinfo  {journal} {Sci
  Rep}\ }\textbf {\bibinfo {volume} {13}} (\bibinfo {year} {2023})}\BibitemShut
  {NoStop}%
\bibitem [{\citenamefont {Xu}\ \emph {et~al.}(2022)\citenamefont {Xu},
  \citenamefont {Özgüler}, \citenamefont {Guglielmo}, \citenamefont {Tran},
  \citenamefont {andLuca Carloni},\ and\ \citenamefont {Fahim}}]{j}%
  \BibitemOpen
  \bibfield  {author} {\bibinfo {author} {\bibfnamefont {D.}~\bibnamefont
  {Xu}}, \bibinfo {author} {\bibfnamefont {A.~B.}\ \bibnamefont {Özgüler}},
  \bibinfo {author} {\bibfnamefont {G.~D.}\ \bibnamefont {Guglielmo}}, \bibinfo
  {author} {\bibfnamefont {N.}~\bibnamefont {Tran}}, \bibinfo {author}
  {\bibfnamefont {G.~N.~P.}\ \bibnamefont {andLuca Carloni}},\ and\ \bibinfo
  {author} {\bibfnamefont {F.}~\bibnamefont {Fahim}},\ }\bibfield  {title}
  {\bibinfo {title} {Neural network accelerator for quantum control},\
  }\href@noop {} {\bibfield  {journal} {\bibinfo  {journal} {IEEE/ACM Third
  International Workshop on Quantum Computing Software (QCS)}\ ,\ \bibinfo
  {pages} {43}} (\bibinfo {year} {2022})}\BibitemShut {NoStop}%
\bibitem [{\citenamefont {Skinner}\ \emph {et~al.}(2006)\citenamefont
  {Skinner}, \citenamefont {Kobzar}, \citenamefont {Luy}, \citenamefont
  {Bendall}, \citenamefont {Bermel}, \citenamefont {Khaneja},\ and\
  \citenamefont {Glaser}}]{rr}%
  \BibitemOpen
  \bibfield  {author} {\bibinfo {author} {\bibfnamefont {T.~E.}\ \bibnamefont
  {Skinner}}, \bibinfo {author} {\bibfnamefont {K.}~\bibnamefont {Kobzar}},
  \bibinfo {author} {\bibfnamefont {B.}~\bibnamefont {Luy}}, \bibinfo {author}
  {\bibfnamefont {M.~R.}\ \bibnamefont {Bendall}}, \bibinfo {author}
  {\bibfnamefont {W.}~\bibnamefont {Bermel}}, \bibinfo {author} {\bibfnamefont
  {N.}~\bibnamefont {Khaneja}},\ and\ \bibinfo {author} {\bibfnamefont {S.~J.}\
  \bibnamefont {Glaser}},\ }\bibfield  {title} {\bibinfo {title} {Optimal
  control design of constant amplitude phase-modulated pulses: Application to
  calibration-free broadband excitation},\ }\href@noop {} {\bibfield  {journal}
  {\bibinfo  {journal} {Journal of Magnetic Resonance}\ }\textbf {\bibinfo
  {volume} {179}},\ \bibinfo {pages} {241} (\bibinfo {year}
  {2006})}\BibitemShut {NoStop}%
\bibitem [{\citenamefont {Kuzmanovi\ifmmode~\acute{c}\else \'{c}\fi{}}\ \emph
  {et~al.}(2024)\citenamefont {Kuzmanovi\ifmmode~\acute{c}\else \'{c}\fi{}},
  \citenamefont {Bj\"orkman}, \citenamefont {McCord}, \citenamefont {Dogra},\
  and\ \citenamefont {Paraoanu}}]{ss}%
  \BibitemOpen
  \bibfield  {author} {\bibinfo {author} {\bibfnamefont {M.}~\bibnamefont
  {Kuzmanovi\ifmmode~\acute{c}\else \'{c}\fi{}}}, \bibinfo {author}
  {\bibfnamefont {I.}~\bibnamefont {Bj\"orkman}}, \bibinfo {author}
  {\bibfnamefont {J.~J.}\ \bibnamefont {McCord}}, \bibinfo {author}
  {\bibfnamefont {S.}~\bibnamefont {Dogra}},\ and\ \bibinfo {author}
  {\bibfnamefont {G.~S.}\ \bibnamefont {Paraoanu}},\ }\bibfield  {title}
  {\bibinfo {title} {High-fidelity robust qubit control by phase-modulated
  pulses},\ }\href@noop {} {\bibfield  {journal} {\bibinfo  {journal} {Phys.
  Rev. Res.}\ }\textbf {\bibinfo {volume} {6}},\ \bibinfo {pages} {013188}
  (\bibinfo {year} {2024})}\BibitemShut {NoStop}%
\bibitem [{\citenamefont {Lu}\ \emph {et~al.}(2016)\citenamefont {Lu},
  \citenamefont {Brodutch}, \citenamefont {Park}, \citenamefont {Katiyar},
  \citenamefont {Jochym-O’Connor},\ and\ \citenamefont {Laflamme}}]{ee}%
  \BibitemOpen
  \bibfield  {author} {\bibinfo {author} {\bibfnamefont {D.}~\bibnamefont
  {Lu}}, \bibinfo {author} {\bibfnamefont {A.}~\bibnamefont {Brodutch}},
  \bibinfo {author} {\bibfnamefont {J.}~\bibnamefont {Park}}, \bibinfo {author}
  {\bibfnamefont {H.}~\bibnamefont {Katiyar}}, \bibinfo {author} {\bibfnamefont
  {T.}~\bibnamefont {Jochym-O’Connor}},\ and\ \bibinfo {author}
  {\bibfnamefont {R.}~\bibnamefont {Laflamme}},\ }\bibfield  {title} {\bibinfo
  {title} {Nmr quantum information processing},\ }\href@noop {} {\bibfield
  {journal} {\bibinfo  {journal} {Electron Spin Resonance (ESR) Based Quantum
  Computing}\ } (\bibinfo {year} {2016})}\BibitemShut {NoStop}%
\bibitem [{\citenamefont {Bhole}(2020)}]{ff}%
  \BibitemOpen
  \bibfield  {author} {\bibinfo {author} {\bibfnamefont {G.}~\bibnamefont
  {Bhole}},\ }\bibfield  {title} {\bibinfo {title} {Coherent control for
  quantum information processing},\ }\href@noop {} {\bibfield  {journal}
  {\bibinfo  {journal} {PhD thesis}\ } (\bibinfo {year} {2020})}\BibitemShut
  {NoStop}%
\bibitem [{\citenamefont {Kingma}\ and\ \citenamefont {Ba}(2014)}]{ttt}%
  \BibitemOpen
  \bibfield  {author} {\bibinfo {author} {\bibfnamefont {D.~P.}\ \bibnamefont
  {Kingma}}\ and\ \bibinfo {author} {\bibfnamefont {J.}~\bibnamefont {Ba}},\
  }\bibfield  {title} {\bibinfo {title} {Adam: A method for stochastic
  optimization},\ }\href {https://api.semanticscholar.org/CorpusID:6628106}
  {\bibfield  {journal} {\bibinfo  {journal} {CoRR}\ }\textbf {\bibinfo
  {volume} {abs/1412.6980}} (\bibinfo {year} {2014})}\BibitemShut {NoStop}%
\bibitem [{\citenamefont {Goodfellow}\ \emph {et~al.}(2016)\citenamefont
  {Goodfellow}, \citenamefont {Bengio},\ and\ \citenamefont {Courville}}]{Gdf}%
  \BibitemOpen
  \bibfield  {author} {\bibinfo {author} {\bibfnamefont {I.}~\bibnamefont
  {Goodfellow}}, \bibinfo {author} {\bibfnamefont {Y.}~\bibnamefont {Bengio}},\
  and\ \bibinfo {author} {\bibfnamefont {A.}~\bibnamefont {Courville}},\
  }\href@noop {} {\emph {\bibinfo {title} {Deep Learning}}}\ (\bibinfo
  {publisher} {MIT Press},\ \bibinfo {year} {2016})\BibitemShut {NoStop}%
\bibitem [{\citenamefont {Singh}\ \emph {et~al.}(2016)\citenamefont {Singh},
  \citenamefont {Arvind},\ and\ \citenamefont {Dorai}}]{gg}%
  \BibitemOpen
  \bibfield  {author} {\bibinfo {author} {\bibfnamefont {H.}~\bibnamefont
  {Singh}}, \bibinfo {author} {\bibnamefont {Arvind}},\ and\ \bibinfo {author}
  {\bibfnamefont {K.}~\bibnamefont {Dorai}},\ }\bibfield  {title} {\bibinfo
  {title} {Constructing valid density matrices on an nmr quantum information
  processor via maximum likelihood estimation},\ }\href@noop {} {\bibfield
  {journal} {\bibinfo  {journal} {Physics Letter A}\ } (\bibinfo {year}
  {2016})}\BibitemShut {NoStop}%
\bibitem [{\citenamefont {Dogra}\ \emph {et~al.}(2015)\citenamefont {Dogra},
  \citenamefont {Dorai},\ and\ \citenamefont {Arvind}}]{dogra2015experimental}%
  \BibitemOpen
  \bibfield  {author} {\bibinfo {author} {\bibfnamefont {S.}~\bibnamefont
  {Dogra}}, \bibinfo {author} {\bibfnamefont {K.}~\bibnamefont {Dorai}},\ and\
  \bibinfo {author} {\bibnamefont {Arvind}},\ }\bibfield  {title} {\bibinfo
  {title} {Experimental construction of generic three-qubit states and their
  reconstruction from two-party reduced states on an nmr quantum information
  processor},\ }\href@noop {} {\bibfield  {journal} {\bibinfo  {journal}
  {Physical Review A}\ }\textbf {\bibinfo {volume} {91}},\ \bibinfo {pages}
  {022312} (\bibinfo {year} {2015})}\BibitemShut {NoStop}%
\bibitem [{\citenamefont {David G.~Cory}(1997)}]{uuuu}%
  \BibitemOpen
  \bibfield  {author} {\bibinfo {author} {\bibfnamefont {T.~F.~H.}\
  \bibnamefont {David G.~Cory}, \bibfnamefont {Amr F.~Fahmy}},\ }\bibfield
  {title} {\bibinfo {title} {Ensemble quantum computing by nmr spectroscopy},\
  }\href@noop {} {\bibfield  {journal} {\bibinfo  {journal} {Proceedings of the
  National Academy of Sciences}\ }\textbf {\bibinfo {volume} {94}},\ \bibinfo
  {pages} {1634} (\bibinfo {year} {1997})}\BibitemShut {NoStop}%
\bibitem [{\citenamefont {Li}\ \emph {et~al.}(2016)\citenamefont {Li},
  \citenamefont {Lu}, \citenamefont {Luo}, \citenamefont {Laflamme},
  \citenamefont {Peng},\ and\ \citenamefont {Du}}]{wwww}%
  \BibitemOpen
  \bibfield  {author} {\bibinfo {author} {\bibfnamefont {J.}~\bibnamefont
  {Li}}, \bibinfo {author} {\bibfnamefont {D.}~\bibnamefont {Lu}}, \bibinfo
  {author} {\bibfnamefont {Z.}~\bibnamefont {Luo}}, \bibinfo {author}
  {\bibfnamefont {R.}~\bibnamefont {Laflamme}}, \bibinfo {author}
  {\bibfnamefont {X.}~\bibnamefont {Peng}},\ and\ \bibinfo {author}
  {\bibfnamefont {J.}~\bibnamefont {Du}},\ }\bibfield  {title} {\bibinfo
  {title} {Approximation of reachable sets for coherently controlled open
  quantum systems: Application to quantum state engineering},\ }\href@noop {}
  {\bibfield  {journal} {\bibinfo  {journal} {Phys. Rev. A}\ }\textbf {\bibinfo
  {volume} {94}},\ \bibinfo {pages} {012312} (\bibinfo {year}
  {2016})}\BibitemShut {NoStop}%
\bibitem [{\citenamefont {Feng}\ \emph {et~al.}(2022)\citenamefont {Feng},
  \citenamefont {Hou}, \citenamefont {Zou}, \citenamefont {Shi}, \citenamefont
  {Yu}, \citenamefont {Sheng}, \citenamefont {Rao}, \citenamefont {Ma},
  \citenamefont {Chen}, \citenamefont {Ren}, \citenamefont {Miao},
  \citenamefont {Xiang},\ and\ \citenamefont {Zeng}}]{vvvv}%
  \BibitemOpen
  \bibfield  {author} {\bibinfo {author} {\bibfnamefont {G.}~\bibnamefont
  {Feng}}, \bibinfo {author} {\bibfnamefont {S.-Y.}\ \bibnamefont {Hou}},
  \bibinfo {author} {\bibfnamefont {H.}~\bibnamefont {Zou}}, \bibinfo {author}
  {\bibfnamefont {W.}~\bibnamefont {Shi}}, \bibinfo {author} {\bibfnamefont
  {S.}~\bibnamefont {Yu}}, \bibinfo {author} {\bibfnamefont {Z.}~\bibnamefont
  {Sheng}}, \bibinfo {author} {\bibfnamefont {X.}~\bibnamefont {Rao}}, \bibinfo
  {author} {\bibfnamefont {K.}~\bibnamefont {Ma}}, \bibinfo {author}
  {\bibfnamefont {C.}~\bibnamefont {Chen}}, \bibinfo {author} {\bibfnamefont
  {B.}~\bibnamefont {Ren}}, \bibinfo {author} {\bibfnamefont {G.}~\bibnamefont
  {Miao}}, \bibinfo {author} {\bibfnamefont {J.}~\bibnamefont {Xiang}},\ and\
  \bibinfo {author} {\bibfnamefont {B.}~\bibnamefont {Zeng}},\ }\bibfield
  {title} {\bibinfo {title} {Spinq triangulum: A commercial three-qubit desktop
  quantum computer},\ }\href@noop {} {\bibfield  {journal} {\bibinfo  {journal}
  {IEEE Nanotechnology Magazine}\ }\textbf {\bibinfo {volume} {16}},\ \bibinfo
  {pages} {20} (\bibinfo {year} {2022})}\BibitemShut {NoStop}%
\bibitem [{xgg(2002)}]{xgg}%
  \BibitemOpen
  \bibfield  {title} {\bibinfo {title} {The quantum state tomography on an nmr
  system},\ }\href@noop {} {\bibfield  {journal} {\bibinfo  {journal} {Physics
  Letters A}\ }\textbf {\bibinfo {volume} {305}},\ \bibinfo {pages} {349}
  (\bibinfo {year} {2002})}\BibitemShut {NoStop}%
\end{thebibliography}%
\end{document}